\documentclass[prb,aps,twocolumn,groupedaddress,floats,final,superscriptaddress,nopacks]{revtex4-1}
\usepackage{graphicx}
\usepackage{subfigure}
\usepackage{amssymb}
\usepackage{latexsym}
\usepackage{epsfig}
\usepackage{psfrag}
\usepackage{dcolumn}
\usepackage{amsmath,amsfonts,amssymb,bm,ulem}
\usepackage{color}
\usepackage{float}

\newcommand{\be}{\begin{equation}}
\newcommand{\ee}{\end{equation}}
\newcommand{\bea}{\begin{eqnarray}}
\newcommand{\eea}{\end{eqnarray}}

\newcommand{\s}{\sigma}

\newcommand{\la}{\langle}
\newcommand{\ra}{\rangle}
\newcommand{\rd}{\mbox{d}}

\begin{document}
\title{Signatures of multiple charge excitations in RIXS spectra of metals}

\author{I. S. Tupitsyn}
\affiliation{Department of Physics, University of Massachusetts, Amherst, MA 01003, USA}
\author{A. M. Tsvelik}
\affiliation{Condensed Matter Physics and Materials Science Division, Brookhaven National Laboratory, Upton, NY 11973-5000, USA}
\author{R. M. Konik}
\affiliation{Condensed Matter Physics and Materials Science Division, Brookhaven National Laboratory, Upton, NY 11973-5000, USA}
\author{N. V. Prokof'ev}
\affiliation{Department of Physics, University of Massachusetts, Amherst, MA 01003, USA}

%\date{\today}

\begin{abstract}
We study how multiple charge excitations appear in the resonant inelastic x-ray scattering (RIXS) spectra of metals. Typically the focus is upon single excitations in the problem, the plasmons and electron-hole pairs, while the multi-excitation processes are neglected. However, at small momentum transfer the multi-excitation contributions may dominate the signal and one needs to understand how to interpret the data in such cases. In particular, we demonstrate how to ''decode" the total multi-excitation intensity and extract the plasmon dispersion. While our calculations are based on the random phase approximation for jellium, which does not allow to obtain quantitatively precise results in the entire region of parameters, we expect them to capture semi-qualitatively all features expected for charged Fermi-liquid states, including universal and singular properties of the RIXS spectra.
\end{abstract}

\maketitle

%%%%%%%%%%%%%%%%%%%%%%%%%%%%%%%%%%%%%%%%%%%%%%%%%%%%%%%%%%%%%%%%%%%%%%%%%%%%%%%%%%%%%%%%%%%%%%%%%
\section{Introduction}
%%%%%%%%%%%%%%%%%%%%%%%%%%%%%%%%%%%%%%%%%%%%%%%%%%%%%%%%%%%%%%%%%%%%%%%%%%%%%%%%%%%%%%%%%%%%%%%%%

Resonant inelastic X-ray scattering (RIXS) is a very promising technique for studying collective excitations in condensed matter systems. \cite{Kotani2001,review,sawat2,chen,sawat4,dean,deanNature,NatureTD} However, extracting the necessary information from the RIXS data is a highly non-trivial task. The deep core hole, created by an incident photon, disturbs the system which leads to a variety of multiparticle processes contributing to the measured RIXS signal. To extract the relevant information about excitations of interest, one needs to properly understand all the processes involved and be able to distinguish single- and multi-excitation processes.

Now traditionally in condensed matter, one is interested in coherent excitations and pays much less
attention to multi-particle continua. In RIXS experiments, however, the single-particle contribution can be weak and masked by multi-particle ones. This turns out to be the case in Coulomb systems where the single-particle contributions (plasmons and $e-h$ pairs) are suppressed at small momentum transfer. \cite{TKPT2019} In that case we have to learn how to extract information about single excitations from multi-excitation processes. These processes (typically characterized by rather broad frequency signals) also contain information about coherent excitations but in a convoluted form.
Thus, understanding their properties is very important, especially in regions of phase space where they dominate the RIXS intensity. In general, studies of multiple excitations may prove to be a convenient practical way of extracting the underlying physics from the broad frequency signal provided its universal features and the corresponding singularities are well understood.

One example of the importance of multi-excitation processes in understanding the RIXS amplitude is the role that bimagnons play in the RIXS response of magnetic materials. The bimagnon response is associated with a four-spin correlation function \cite{Brink2007,klauser} (whereas the magnonic excitation is probed by a more standard two point spin-correlation function). It has been demonstrated that the bimagnon signal forms an important part of the RIXS signal.\cite{Brink2007,Igarashi2007,Chubuk2007,Vernay2007,Forte2008} (The bimagnon response also plays a key role in magnetic Raman scattering, see for instance Refs.~\onlinecite{Tsutsui1999,Harada2002,TD2010}.)

In this work, in a similar way, we want to understand the role that the multi-excitation processes play in metallic systems where a correct treatment of dynamic screening of long-range Coulomb interactions is necessary. Our goal is ultimately to learn how to ``decode" the total multi-excitation intensity and extract all the important information from it.

There are several approaches for calculating the Kramers-Heisenberg amplitude that describes RIXS spectra. These include single particle approaches \cite{Johnson,Carlisle,Jia} appropriate for weakly correlated systems, methods that emphasize the excitonic state formed by the core hole and the excited electron,\cite{veenendaal,shirley} non-perturbative approaches such as exact diagonalization,\cite{Ghiringhelli,Kotani2001,Hancock,EDRIXS} dynamical mean field theory,\cite{DMFT} integrability,\cite{klauser} and the density matrix renormalization group,\cite{DMRG} methodologies that emphasize the idea that the Anderson orthogonality catastrophe is at the heart of the RIXS response, \cite{Ahn,rehr,demler1,demler2} and finally diagrammatic methods.\cite{Nomura,Platisa,phononsTD,Markiewicz,TKPT2019} Because we are interested in understanding the multi-excitation response, it is natural to employ the latter.  The  advantage of the diagrammatic approach is that one can analytically isolate the effects of the different multi-excitation branches in the presence of long-ranged Coulomb interactions.  Hence we will follow the diagrammatic framework laid out in our earlier work, Ref.~\onlinecite{TKPT2019}.

As explained in Ref.~\onlinecite{TKPT2019} and discussed briefly in Section \ref{FM}, we work in the ultra-short core hole lifetime limit.  This gives as a natural small parameter $\Omega_{pl}/\Gamma$ where $\Omega_{pl}$ is the plasmon frequency and $\Gamma^{-1}$ is the core hole lifetime.  The natural material candidates for our work will then be Dirac and topological metals involving first row transition metal elements (i.e. FeTe$_{1-x}$Se$_x$ family) whose K-edges are accessible on current generation hard x-ray RIXS beamlines.
%{\color{red} An additional advantage for working with these materials is that the low energy sectors frequently possess rotational invariance which greatly simplifies the calculations.}

An outline of the paper is as follows. In Section \ref{FM} we provide a description of our model and the attendant formalism used to solve it for the indirect RIXS response. In Sections \ref{SEP} and \ref{ME} we discuss the contribution of single- and multi-excitations to the RIXS response respectively. In Section \ref{PD} we use the discussion of the previous two Sections to provide a
derivation of the dispersion of the plasmon from the two-excitation contribution to the RIXS signal.   This is one of the main results of the paper. In Section \ref{TOT} we consider in additional detail how the single- and multi-particle spectra can be seen in measurements of the RIXS signal in metals.  Finally in Section \ref{CON} we offer our conclusions.

%%%%%%%%%%%%%%%%%%%%%%%%%%%%%%%%%%%%%%%%%%%%%%%%%%%%%%%%%%%%%%%%%%%%%%%%%%%%%%%%%%%%%%%%%%%%%%%%%
\section{Model and Formalism for Indirect RIXS}
\label{FM}
%%%%%%%%%%%%%%%%%%%%%%%%%%%%%%%%%%%%%%%%%%%%%%%%%%%%%%%%%%%%%%%%%%%%%%%%%%%%%%%%%%%%%%%%%%%%%%%%%

\begin{figure}[tbh]
\centerline{\includegraphics[angle = 0,width=0.9\columnwidth]{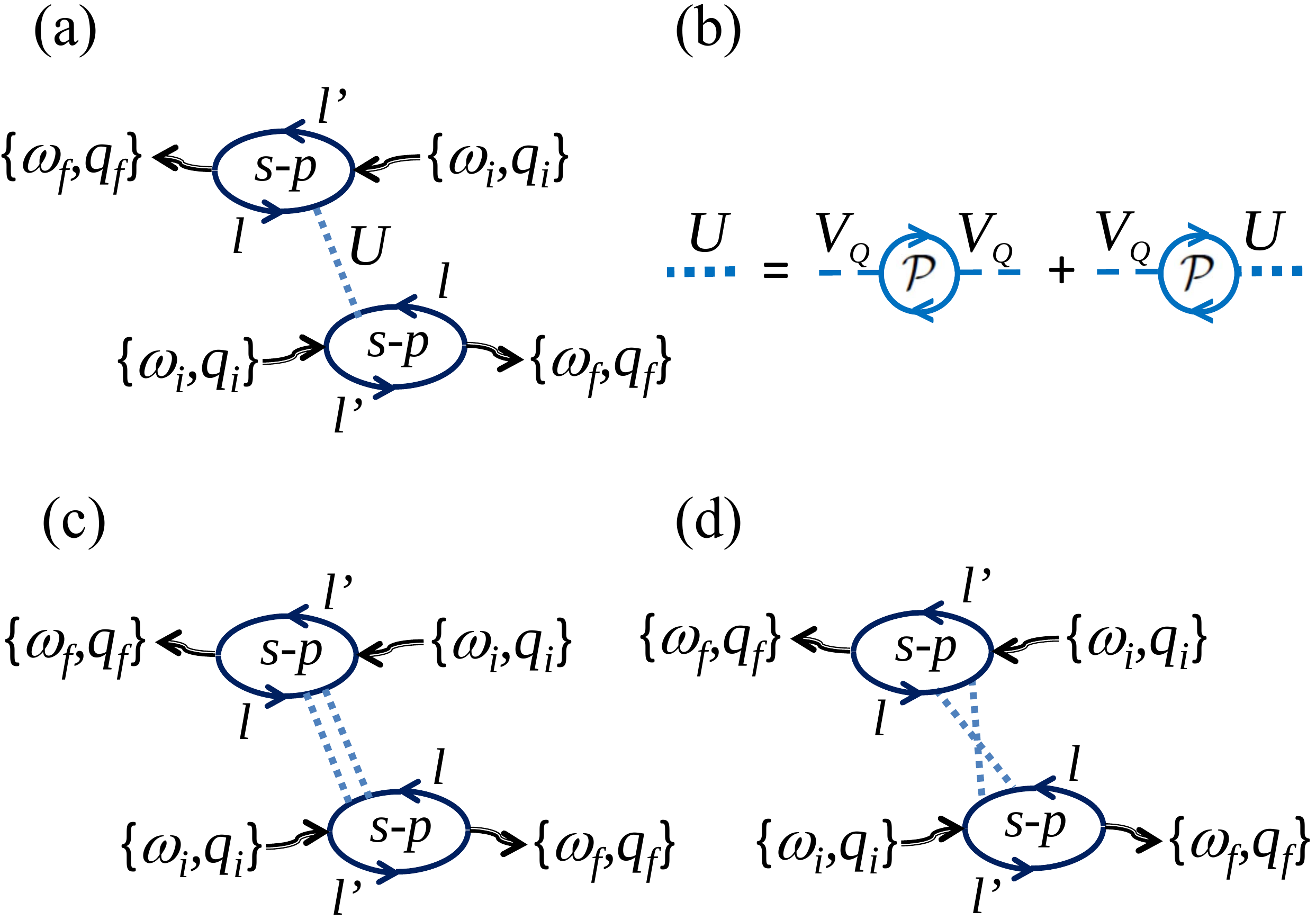}}
\caption{(color online) The leading order diagrams for the RIXS cross section in the UCL limit (e.g., for the $K$-edge case) with $l$ and $l'$ indexes standing for either $s$-hole, or $p$-electron, and $\{\omega_i,{\bf q}_i\}$ and $\{\omega_f,{\bf q}_f\}$ specifying the incoming and outgoing photon energies and momenta. The dynamically screened Coulomb potential $U$, shown by the
blue dotted line, can connect any upper $s-p$ bubble line ($s$ or $p$) with any lower $s-p$ bubble line; here, for brevity, we present only one particular way of connecting the bubbles. }
\label{Fig01}
\end{figure}

In the indirect RIXS process, the incident photon with energy $\omega_i$ and momentum ${\bf q_i}$ excites the deep core electron into the high-energy (potentially mobile) $p$ state and leaves behind a (localized) core hole $s$. [Our labels for bands are arbitrary and can change depending on the material studied.] Eventually, the $p$-electron repopulates the $s$-hole through the emission of a photon with energy $\omega_f$ and momentum ${\bf q_f}$. In what follows we are going to work in the ultra-short core hole life time limit (UCL) \cite{Brink2005,Brink2006,Ament2007,TKPT2019}, frequently employed in the RIXS calculations. In this limit the $s-p$ dipole emits low-energy collective excitations (represented by bosonic lines $U$ in Fig.~\ref{Fig01}) only during the very short time interval, $\Gamma^{-1}$. The differences, ${\bf Q} = {\bf q}_i - {\bf q_f}$ and $\Omega = \omega_i - \omega_f$, describe the corresponding momentum and energy transfer to excitations.

The Hamiltonian appropriate for the computation of indirect RIXS intensity has the following description (see, for instance, Ref. \onlinecite{shvaika}).  To formulate the Feynman diagram expansion for the RIXS cross section one needs  to introduce two replicas of core electrons (with creation operators $s^\dagger_{a}, a =1,2$) localized at different space points at a distance ${\bf R}_{12}$ from each other, and  two {\color{red} replicas} of high energy electrons (created by $p^\dagger_a$), and a species of conduction electrons (created by $d^\dagger$).  The different replicas enter into different $s-p$ fermion bubbles depicted on Fig. 1(a,c,d). The corresponding Anderson model can be formulated as follows:
\bea
H &=& H_s + H_d + H_p + H_{dd}  + H_{sp} + H_{sd} + H_{pd} ;   \\
H_s &=& \sum_{\s, a=1,2} \epsilon_s \; s^{\dagger}_{\s,a} \; s^{\;}_{\s,a} + H_{s,\Gamma}; \nonumber\\
H_p &=& \! \sum_{k, \s, a} \epsilon_p({\bf k}) \; p^{\dagger}_{{\bf k},m\s,a } \; p^{\;}_{{\bf k},m\s,a }; \;
H_d =   \! \sum_{k \s} \epsilon_d({\bf k}) \; d^{\dagger}_{{\bf k},\s} \; d^{\;}_{{\bf k},\s};  \nonumber \\
H_{sd} &=& -\int d{\bf r} \, n_d({\bf r})\times \Big[\frac{e^2}{|{\bf R}_{1}- {\bf r}|} \, n_{s,1}
+ \frac{e^2}{|{\bf R}_{2} - {\bf r}|} \, n_{s,2}  \Big],
\nonumber
\label{anderson}
\eea
where $\sigma=\pm$ is the spin index, $\epsilon_s$ gives the energy of the core electron, $\epsilon_p({\bf k})$/$\epsilon_d({\bf k})$ gives the dispersion of the $p/d$ electrons,
and $n_{s,d,p}$ are the corresponding number densities. $H_{s,\Gamma}$ defines the core hole with a finite lifetime $\Gamma^{-1}$. The interaction Hamiltonians $H_{sp}$, $H_{pd}$, and $H_{dd}$ have similar structures based on the Coulomb potential, $V_{{\bf r}} = e^2/r$ or  $V_{{\bf Q}} = 4\pi e^2/Q^2$ (for brevity, we present explicitly only $H_{sd}$). Here and below we are using units such that the Planck's constant, $\hbar$, the Fermi momentum, $k_F$, and the Fermi energy, $\epsilon_F$, are set to unity.

By integrating the $d$-electrons out, one arrives at the model where $s$-holes/$p$-electrons are coupled by the action
\bea
&&S = S_0 + \int \rd \tau_1\rd \tau_2 d{\bf r}_1 d{\bf r}_2
\rho({\bf r}_1,\tau_1)U( {\bf r}_{12}, \tau_{12}) \rho({\bf r}_2, \tau_2), \nonumber \\
&&\rho({\bf r}) = \sum_{a} \left[ \delta({\bf r} - {\bf R}_a) n_{s,a} + n_{p,a} ({\bf r}) \right],
\label{anderson2}
\eea
where $S_0$ is the bare action for $s$-holes/$p$-electrons and $U$ is the dynamic part of the screened Coulomb potential. In this formulation, the correlation function, $\chi_{R}$, responsible for the RIXS cross section, can be written as (see, for instance, Refs. \onlinecite{demler1,demler2})
\bea
&& \chi_{R}({\bf R}_{12}; t_1,t_2,t) = \label{corr} \\
&& \la D_1(q_i,t_1/2) \; D^\dagger_1(q_f,-t_1/2) \; \cr\cr
&& \hskip .5in  D^\dagger_2(q_i,t_2/2 + t) \; D_2(q_f,-t_2/2 + t) \ra_{ret}, \nonumber
\eea
where the dipole operator on site ${\bf R}_a$ is defined as
\begin{equation}
D^\dagger_a(q) = \sum_m P_m(q) \epsilon_m \sum_{k,\s}p_{k+q,m\s,a} s^\dagger_{\s, a}
\end{equation}
The RIXS cross section is then extracted from the imaginary part of the analytically continued Fourier transform of this correlation function in direct analogy to the Raman scattering response. \cite{shvaika}
Here we have made explicit the polarization of the incident photon: $\epsilon_m$, $m=\pm 1, 0$ while $P_m(q)$ is an appropriate atomic matrix element dependent on the photon momenta.  We note that because our system is rotationally invariant, we are computing the part of the RIXS response that contains the elastic peak.\cite{review}  This might then obscure low energy features of interest to us.  However the plasmon dispersion, one of the main foci of this paper, is a finite frequency phenomenon.  While the particular polarization used experimentally is relevant for the RIXS response in specific materials, we will simply absorb factors of $\epsilon_m$ into an overall prefactor (below $\Upsilon_i$).  We also note that we will not take into account the q-dependence of the atomic form factor, $P_m(q)$, although in real materials this will effect the overall response, particularly at large $q$.

Within the diagrammatic formalism of Ref.~\onlinecite{TKPT2019} the first order (in the dynamically screened interaction $U$, see Fig.~\ref{Fig01}(b))) contribution to the RIXS intensity is shown in Fig.~\ref{Fig01}(a). It describes the single-excitation contribution from plasmons and $e-h$ pairs. The second order diagrams, Figs.~\ref{Fig01}(c) and (d), are responsible for contributions from the two-excitation processes involving either two plasmons, or two $e-h$ pairs, or one plasmon and one $e-h$ pair (we call the latter a hybrid process). Contributions from higher order processes are small because each additional $U$-line comes with the extra factor $\propto \Gamma^{-2}$.  Such processes would include interactions between single excitations, something that we then do not account for here.

The first-order contribution to the RIXS intensity for momentum and energy/frequency transfers is given by
\be
\chi^{(1)}_{R} =  \frac{\Upsilon^2_1}{\Gamma^2}\,  D^{(1)}(\Omega,Q)
\label{Chi1}
\ee
with
\be
D^{(1)} = f_1({\bf Q}) \, {\cal I}m U(\Omega, Q),
\label{D1}
\ee
where the dynamically screened Coulomb potential $U$ is given by (see Fig.~\ref{Fig01}(b))

\be
U(\Omega, Q) = \frac{V_Q}{1 - V_Q {\cal P}(\Omega,Q)} - V_Q; \;\;\; V_Q = \frac{4 \pi e^2}{Q^2} \,
\label{FullU}
\ee
and the form-factor, $f_1({\bf Q})$ vanishes in the small momentum transfer limit; \cite{TKPT2019} ${\cal P}$ is the polarization function. For crystals with inversion symmetry it can be modeled by
\be
f_1({\bf Q}) = \left[ \frac{Q^2}{Q^2+Q^2_d} \right]^2 \;.
\label{f1}
\ee
with some characteristic momentum cutoff $Q_d \sim k_F$. In the UCL limit the $(s-p)$ bubbles
contribute a frequency independent factor and one can replace $\Upsilon_1$ in Eq.~\ref{Chi1} with a constant of the order of unity. 
In what follows, we will use $1/k_F$ and $\varepsilon_F$ as units of length and energy and not mention them explicitly in dimensionless ratios $Q/k_F$ and $\Omega/\varepsilon_F$.

 We note that $\Upsilon_1$ and its counterpart $\Upsilon_2$ in Eqn.~14 contain atomic form factors and so encode the lack of rotational invariance present in real materials.  Our focus in this paper is not on the physics contained within these form factors.  We further note that we do not consider our results accurate when the UCL approximation is violated.

An accurate theoretical prediction requires knowledge of the dynamic dielectric response function, $\epsilon({\bf q},\omega)$. This is a difficult computational materials science problem even for single-excitation processes. It becomes more acute for multi-excitation processes. On the other hand, significant qualitative and even semi-quantitative gains in understanding can be obtained by approximating the dielectric response with the analytic form based on the Lindhard function (equivalent to the random phase approximation (RPA)) which captures all the properties characteristic of the charged Fermi-liquid state.

In our previous work, \cite{TKPT2019} we employed a simplified description of the dielectric response function valid in the $q/k_F \ll 1$ limit. The two most important qualitative features not captured by this treatment were the singular behavior at momenta and frequencies corresponding to the end point of the plasmon dispersion relation, and the non-analyticity at $q=2k_F$. Furthermore for the process involving two $(e-h)$ pairs, even at small momentum and energy transfer one cannot ignore pairs with large momenta $\sim k_F$. To eliminate these deficiencies and achieve a comprehensive semi-quantitative description in a broad parameter range, here we consider the full Lindhard function, \cite{Lindhard,Bogdan2011} correctly capturing all the Fermi-surface effects and the interplay between the plasmon mode and the $e-h$ continuum. This gives for ${\cal P}$:
\bea\label{lindhard}
{\cal R}e {\cal P} &=& \frac{3 n}{4 \varepsilon_F} \Big[ -1
+ \frac{4 Q^2 - Q^4_{-}}{8 Q^3} \log \left| \frac{2Q + Q^2_{-}}{2Q-Q^2_{-}} \right| \nonumber \\
&-& \frac{4 Q^2 - Q^4_{+}}{8 Q^3} \log \left| \frac{2Q + Q^2_{+}}{2Q-Q^2_{+}} \right| \Big], \label{Lindh} \\
{\cal I}m {\cal P} &=& \frac{3 \pi n }{ 8 \varepsilon_F Q}
\left\{
\begin{array}{ll}
-  \Omega & \;\;\;\; \hbox{if A;} \\
-  [1 - (\Omega/Q - Q)^2/4] & \;\;\;\; \hbox{if B;} \\
-  [1 - (\Omega/Q - Q)^2/4] & \;\;\;\; \hbox{if C;}
\end{array}
\right. \label{Lindh_Im} \\
A&:&  \;\;\; Q < 2, \;\;\;\;\;\;\;\;\;\;\;\;\;\;\; 0 \; \leq \Omega < -Q^2 + 2Q; \nonumber \\
B&:&  \;\;\; Q < 2, \; -Q^2 + 2 Q \; \leq \Omega \leq \;\; Q^2 + 2Q; \nonumber \\
C&:&  \;\;\; Q \geq 2, \;\;\;\; Q^2 - 2 Q \; \leq \Omega \leq \;\; Q^2 + 2Q, \nonumber
\eea
where $Q^2_{\pm} = \Omega  \pm Q^2$ and $n=k^3_F/3 \pi^2$.
The definition of the Coulomb $r_s$ parameter is standard: $e^2 = (r_s k_F/m) (4/9\pi)^{1/3}$.

\begin{figure}[tbh]
\centerline{\includegraphics[angle = 0,width=0.8\columnwidth]{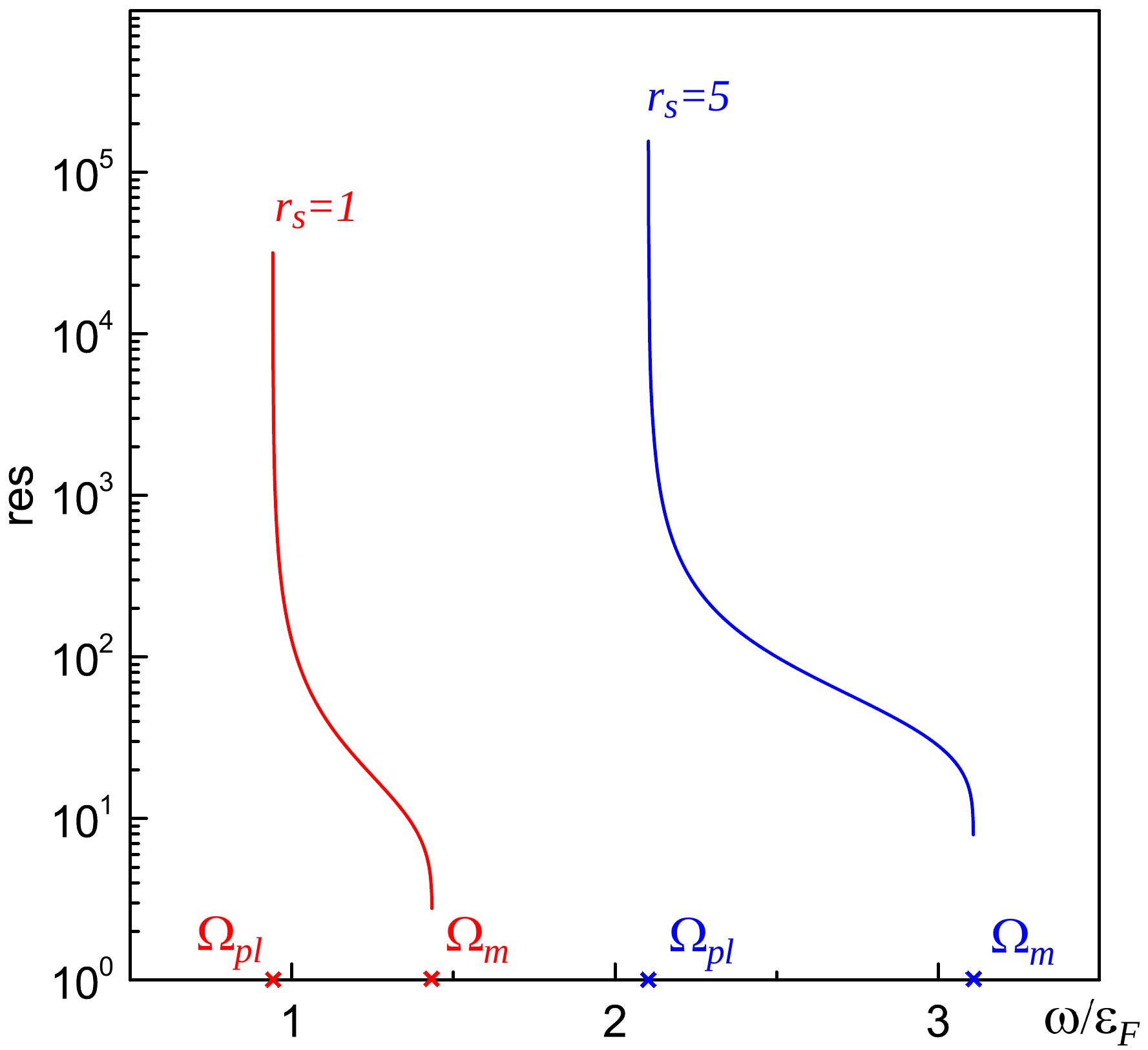}}
\caption{(color online) Plasmon residue $Res(\omega)$ as a function of frequency for $r_s=1$ and $5$. At $\omega \to \Omega_{pl}$, $Res(\omega)$ behaves as $\propto 1/(\omega-\Omega_{pl})$; at
$\omega \to \Omega_m$, $Res(\omega) - Res(\Omega_m) \propto \sqrt{\Omega_m - \omega}$. }
\label{Fig04}
\end{figure}

For momenta $Q<Q_m$, the ${\cal I}m U \equiv \tilde{D} $ function can be naturally divided into two contributions $\tilde{D} = \tilde{D}_{e-h} + \tilde{D}_{pl}$:
the first one is associated with the electron-hole continuum, and the other one with the
gapped plasmon resonance that exists as a separate sharp mode only up to
a finite momentum $Q_m$ when it merges into the $e-h$ continuum. In terms of the real and imaginary parts of the polarization function, we have
\bea
\tilde{D}_{e-h} &=& -\frac{{\cal I}m {\cal P}}{[(V^{-1}_Q - {\cal R}e {\cal P})^2 + ({\cal I}m {\cal P})^2]}\, ,
\label{DEH} \\
\tilde{D}_{pl} &=& \pi \; Res[\omega_{pl}(Q)] \; \delta(\Omega - \omega_{pl}(Q)), \;\;
{\cal I}m {\cal P} = 0. \;\;\;\;\;   \label{DPL}
\eea
The plasmon peak frequency and residue are derived from properties of the dielectric
function $\epsilon = 1 - V_Q  {\cal P}$:
\be
\epsilon (Q, \omega_{pl}(Q))=0; \;\;\;\;
Res[\omega] = \frac{1}{|\partial \epsilon /\partial\Omega|_{\omega}} \,,
\label{RES}
\ee
see Fig.~\ref{Fig04}.

The second-order contribution to the RIXS intensity reads
\be
\chi^{(2)}_{R} =  \frac{\Upsilon^2_2}{\Gamma^4} D^{(2)}(\Omega,Q) \, ,
\label{Chi2}
\ee
with frequency independent constant $\Upsilon_2 \sim 1$ and
\be
D^{(2)}= \int \frac{d{\bf q} d\omega}{(2\pi)^{4}} f_2({\bf q},{\bf Q}-{\bf q})
\tilde{D} (\omega, {\bf q}) \tilde{D} (\Omega -\omega, {\bf Q}-{\bf q})\, ,
\label{D2}
\ee
where the model form-factor
\be
f_2({\bf q}_1,{\bf q}_2) = \frac{ ( {\bf q_1} \cdot {\bf q_2} )^2 }{(q^2_1+Q_d^2)(q^2_2+Q_d^2)}\, .
\label{f2}
\ee
is designed to respect the small momentum transfer dependence coming from the $(s-p)$
bubbles for the sum of two second-order diagrams (see Figs.~\ref{Fig01} (c) and (d)).
Below we set $\Upsilon_{1,2} = 1$ and present all results without the $\Gamma$ factors.

Note that the diagrams in Fig.~\ref{Fig01} do not account for interactions between
excitations---the corresponding processes occur only in higher orders with respect to the
diagrammatic expansion in terms of the number of $U$-lines. Their consideration goes
beyond the scope of present work.

%%%%%%%%%%%%%%%%%%%%%%%%%%%%%%%%%%%%%%%%%%%%%%%%%%%%%%%%%%%%%%%%%%%%%%%%%%%%%%%%%%%%%%%%%%%%%%%%%
\section{Single-excitation processes}
\label{SEP}
%%%%%%%%%%%%%%%%%%%%%%%%%%%%%%%%%%%%%%%%%%%%%%%%%%%%%%%%%%%%%%%%%%%%%%%%%%%%%%%%%%%%%%%%%%%%%%%%%

\begin{figure}[h]
\centerline{\includegraphics[angle = 0,width=0.8\columnwidth]{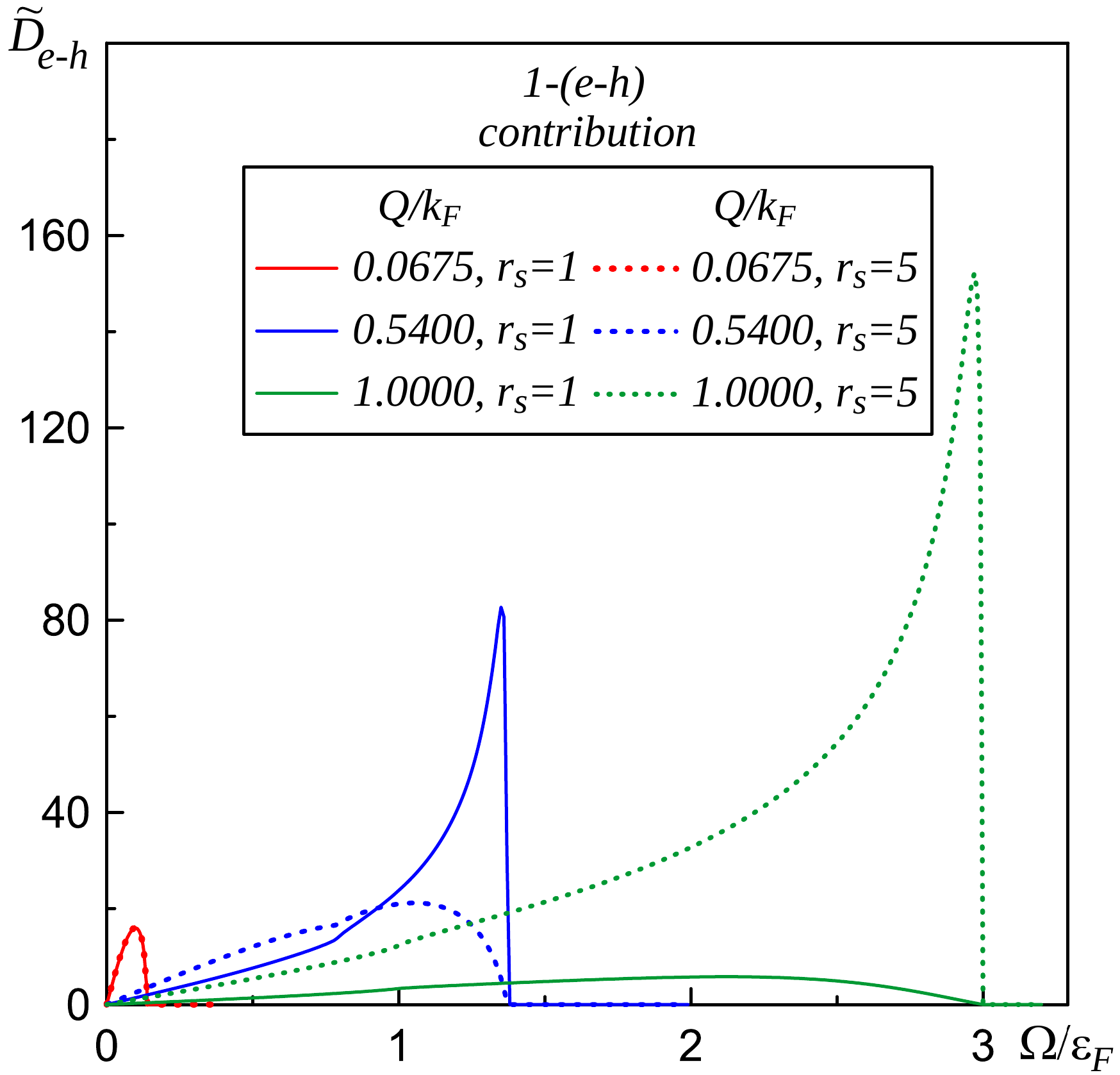}}
\caption{(color online) Single pair contribution to intensity, $\tilde{D}_{e-h}$,
as a function of $\Omega$ for different values of momentum transfer
$Q$ at $r_s=1$ (solid lines) and $5$ (dotted lines). The $f_1(Q)$
factor is removed.}
\label{Fig05}
\end{figure}

Here we briefly review the known properties of $\tilde{D}$. There is no point in showing the plasmon contribution because it is completely characterized by the dispersion relation and the pole residue, see Eq.~(\ref{RES}). %and Figs.~(\ref{Fig02}) and (\ref{Fig04}).
The $e-h$ pair contribution results in a broad spectral curve with intensity diverging on the approach to the end point of the plasmon spectrum, see Fig.~\ref{Fig05}. The upper threshold of the $e-h$ spectrum is located at
\be
\Omega_{e-h}(Q) = v_F Q + \frac{Q^2}{2m} \;.
\label{WEH1}
\ee
The end point of the plasmon dispersion is then defined by the condition
$\Omega_{e-h}(Q_m)=\Omega_m$ leading to the following relation for $\Omega_m$:
\be
\Omega_m = \omega_{pl}(Q_m) = v_F Q_m + \frac{Q_m^2}{2m} \,.
\label{WEH2}
\ee
For $r_s = 1$ and $r_s=5$ the largest plasmon momentum equals to
$Q_m=0.560\, k_F$ and $Q_m=1.027\, k_F$, respectively.

At small momenta $Q$ the shapes of the $\tilde{D}_{e-h}$ curves for different values of $r_s$ are nearly indistinguishable. With increasing $Q$ the peak amplitude increases until $Q=Q_m$, where the $e-h$ continuum ``absorbs'' the plasmon mode. At $Q>Q_m$ the peak maximum decreases while the plasmon contribution no longer exists. For $r_s=1$ the $Q = k_F$ case corresponds to $Q>Q_m$ when the intensity is already rather small, featureless, and broad; for $r_s=5$  this momentum transfer is slightly below the plasmon end-point, $Q<Q_m$, and the intensity keeps increasing in a singular fashion.

\begin{figure}[tbh]
\centerline{\includegraphics[angle = 0,width=0.8\columnwidth]{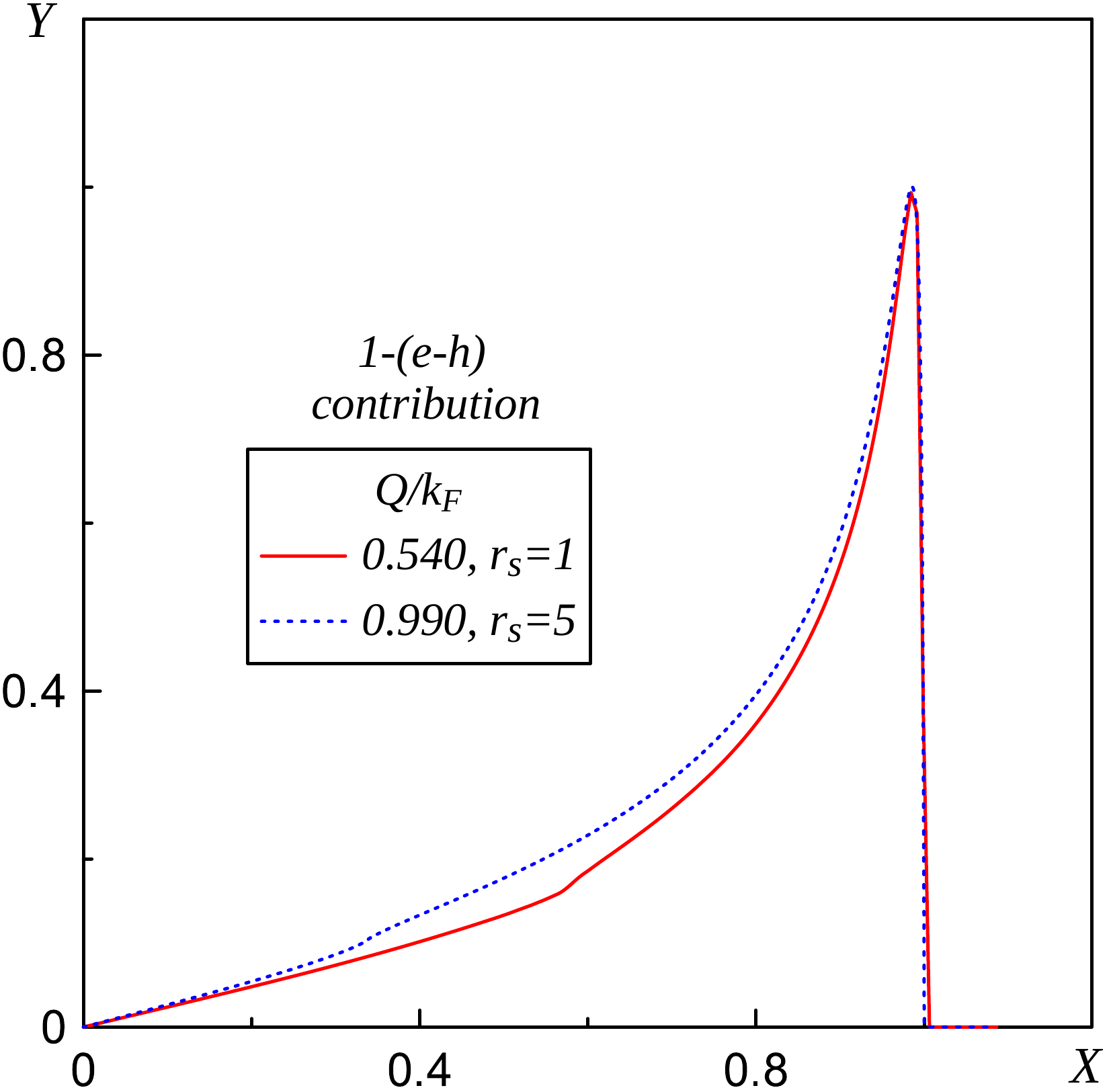}}
\caption{(color online) Scaled single-pair contributions to intensity,
$Y=\tilde{D}_{e-h}/ \{ \tilde{D}_{e-h}\}_{max}$, as functions of $X=\Omega/\Omega_{e-h}(Q)$ for $r_s=1$ and $r_s=5$ (with the same $Q/Q_m \approx 0.964$ ratio for both curves). The $f_1(Q)$ factor is removed.}
\label{Fig06}
\end{figure}

At the RPA level, there is a certain degree of universality in the scaled shapes of the curves for small and large values of $r_s$. It is clear from Fig.~\ref{Fig05} that large values of $r_s$ do not introduce new spectral features; in general, only the positions of peaks and their intensities are changed. To verify this quantitatively, in Fig.~\ref{Fig06} the $e-h$ contributions for $r_s=1$ and $r_s=5$ are presented for momentum transfers corresponding having the same $Q/Q_m$ ratio close to unity. The curves are scaled to have the same peak amplitude and are plotted as functions of the $\Omega/\Omega_m$. It is clear that the characteristic features of the $e-h$ contribution, shown in Fig.~\ref{Fig05} and described above, barely change as a function of $r_s$.

\begin{figure}[htp]
\centering
\subfigure{\includegraphics[angle = 0,width=0.8\columnwidth]{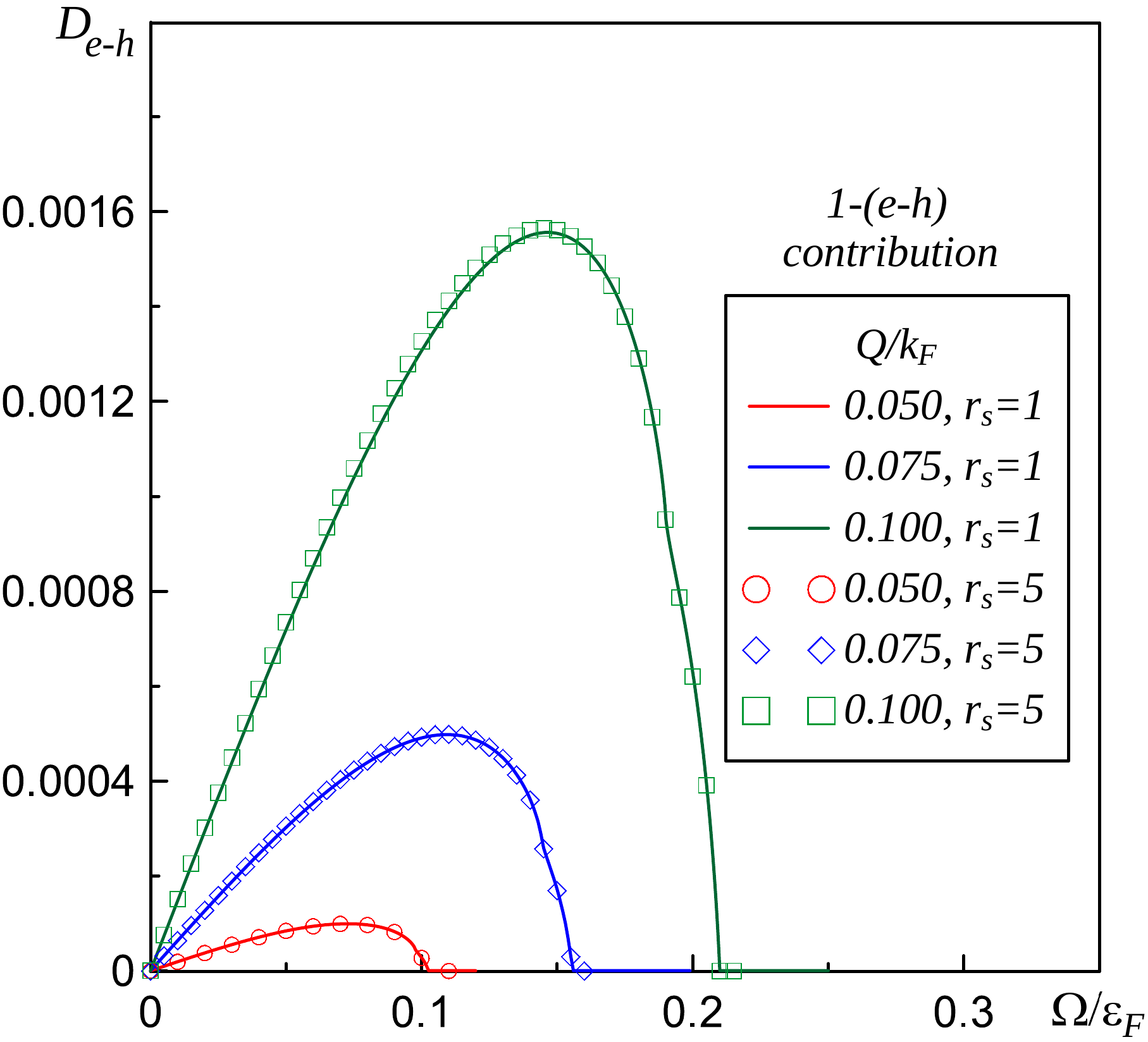}}
\caption{(color online) Single-pair contributions to intensity as functions of $\Omega$ at $r_s=1$ and $5$ when $Q/Q_m <<1$ for both curves. The $f_1(Q)$ factor is not removed, in contrast to Figs. \ref{Fig05} and \ref{Fig06} - this allows to see the actual values of the $e-h$ intensity (without the $\Gamma$ factor). }
\label{Fig07}
\end{figure}

At small momentum (and energy) transfer the $e-h$ intensity curves become independent of $r_s$, see Fig.~\ref{Fig07}. At $\Omega << v_F Q $ they are linear in $\Omega$ and the amplitude is proportional to $Q^3$. The intensity of the plasmon peak under the same conditions is proportional to $r_s^{3/2}Q^2$. These scaling laws imply that at small $Q$ the single-excitation RIXS spectrum can be weak and potentially subdominant to higher-order processes considered next.

%%%%%%%%%%%%%%%%%%%%%%%%%%%%%%%%%%%%%%%%%%%%%%%%%%%%%%%%%%%%%%%%%%%%%%%%%%%%%%%%%%%%%%%%%%%%%%%%%
\section{Multi-excitation processes}
\label{ME}
%%%%%%%%%%%%%%%%%%%%%%%%%%%%%%%%%%%%%%%%%%%%%%%%%%%%%%%%%%%%%%%%%%%%%%%%%%%%%%%%%%%%%%%%%%%%%%%%%

We begin by noting that in the UCL limit the higher-order (with respect to the number of $U$-lines connecting the $s-p$ dipoles) processes are suppressed. However, at small momentum transfer the second-order processes can dominate the spectrum even in the region where the first-order intensity is non-zero because the form factor $f_2$ has a different dependence on $Q$, see Eq.~(\ref{f2}). Simultaneously, as is has been noted in the Introduction, the two-excitation spectra can be used to extract information about single excitations provided the origin of their characteristic features is well understood.

\begin{figure}[h]
\vspace{-2mm}
\subfigure{\includegraphics[scale=0.24]{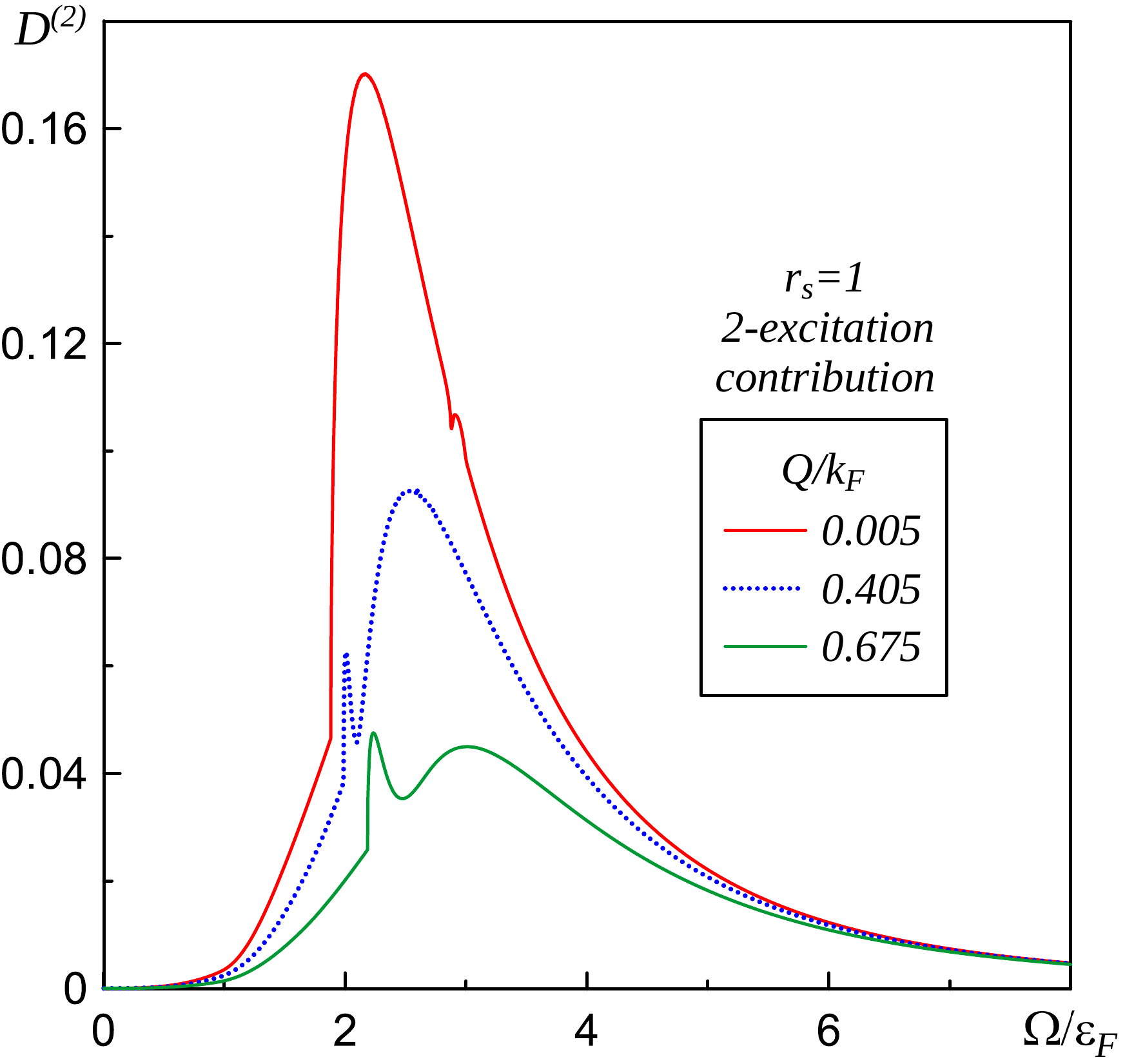}}
\subfigure{\includegraphics[scale=0.24]{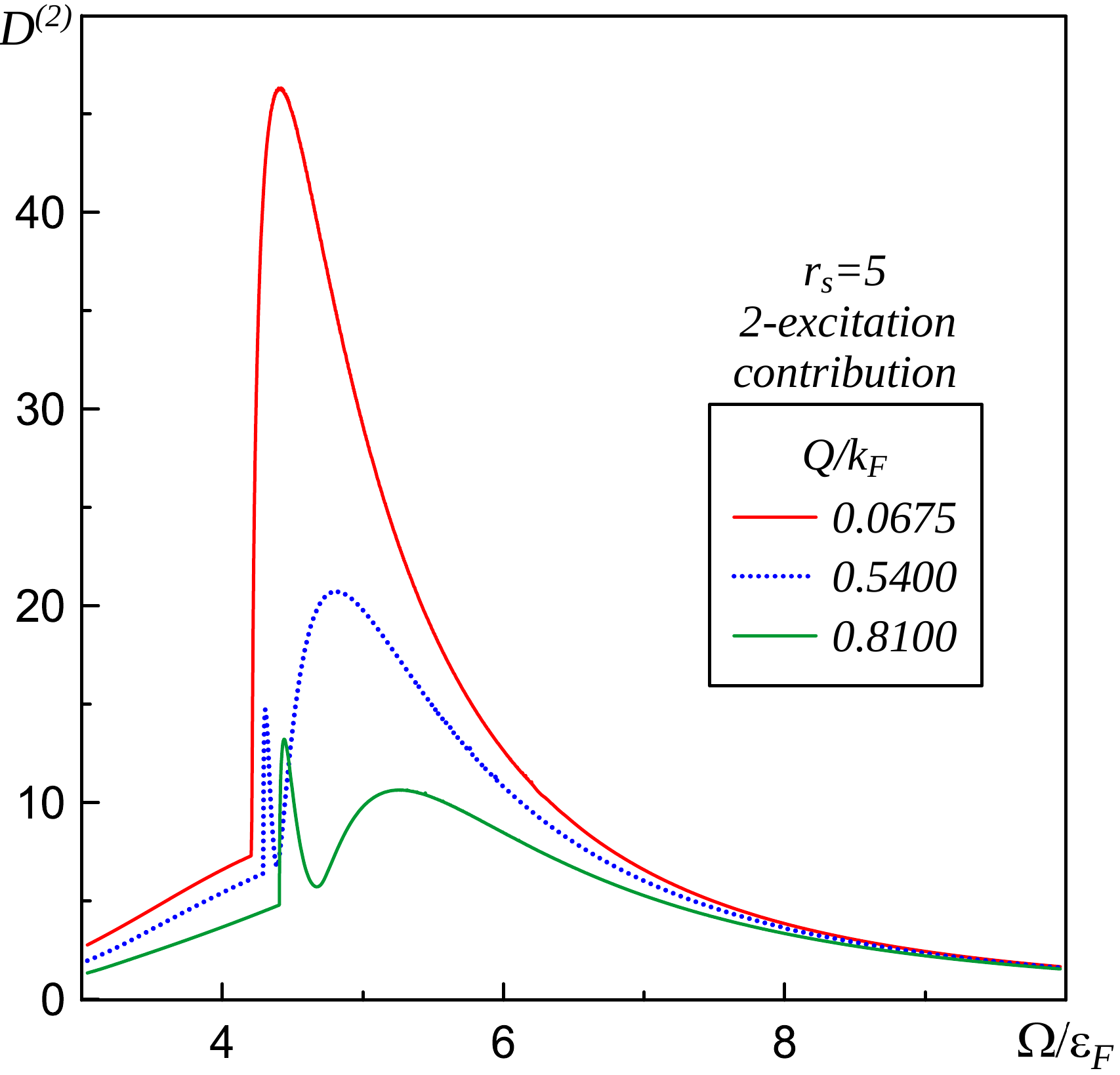}}
\vspace{-3mm}
\caption{(color online) Total two-excitation RIXS intensity as a function of energy transfer $\Omega$. It includes two-plasmon, two-$(e-h)$, and the hybrid (see text) processes shown for several momentum transfers $Q$. Upper panel: $r_s=1$. Lower panel: $r_s=5$. }
\label{Fig07a}
\end{figure}

In Fig.~\ref{Fig07a} we present the example of the entire two-excitation RIXS intensity for two different values of $r_s$. The spectral curves are characterized by some distinct features and in this Section we are going to describe these features in detail.

By substituting equations (\ref{DEH}) and (\ref{DPL}) into equation (\ref{D2}) we obtain partial (two-plasmon, two-($e-h$), and hybrid) contributions to the total intensity. In what follows we first address each process separately and then conclude with the discussion of the total two-excitation intensity.

%%%%%%%%%%%%%%%%%%%%%%%%%%%%%%%%%%%%%%%%%%%%%%%%%%%%%%%%%%%%%%%%%%%%%%%%%%%%%%%%%%%%%%%%%%%%%%%%%
\subsection{Two-plasmon processes}
\label{2PL}
%%%%%%%%%%%%%%%%%%%%%%%%%%%%%%%%%%%%%%%%%%%%%%%%%%%%%%%%%%%%%%%%%%%%%%%%%%%%%%%%%%%%%%%%%%%%%%%%%

In contrast to the sharp single-plasmon peak located at $\omega_{pl}(Q)$, the two-plasmon spectrum is broad - it starts at $2\omega_{pl}(Q/2)$ and terminates at $2\Omega_m$. Even for $Q=0$ the two-plasmon spectrum is well outside of the single-plasmon dispersion range, i.e., the single- and two-plasmon spectra do not overlap. However, single-pair and two-plasmon peaks can overlap at large momentum transfer.

\begin{figure}[htp]
\centering
\subfigure{\includegraphics[scale=0.24]{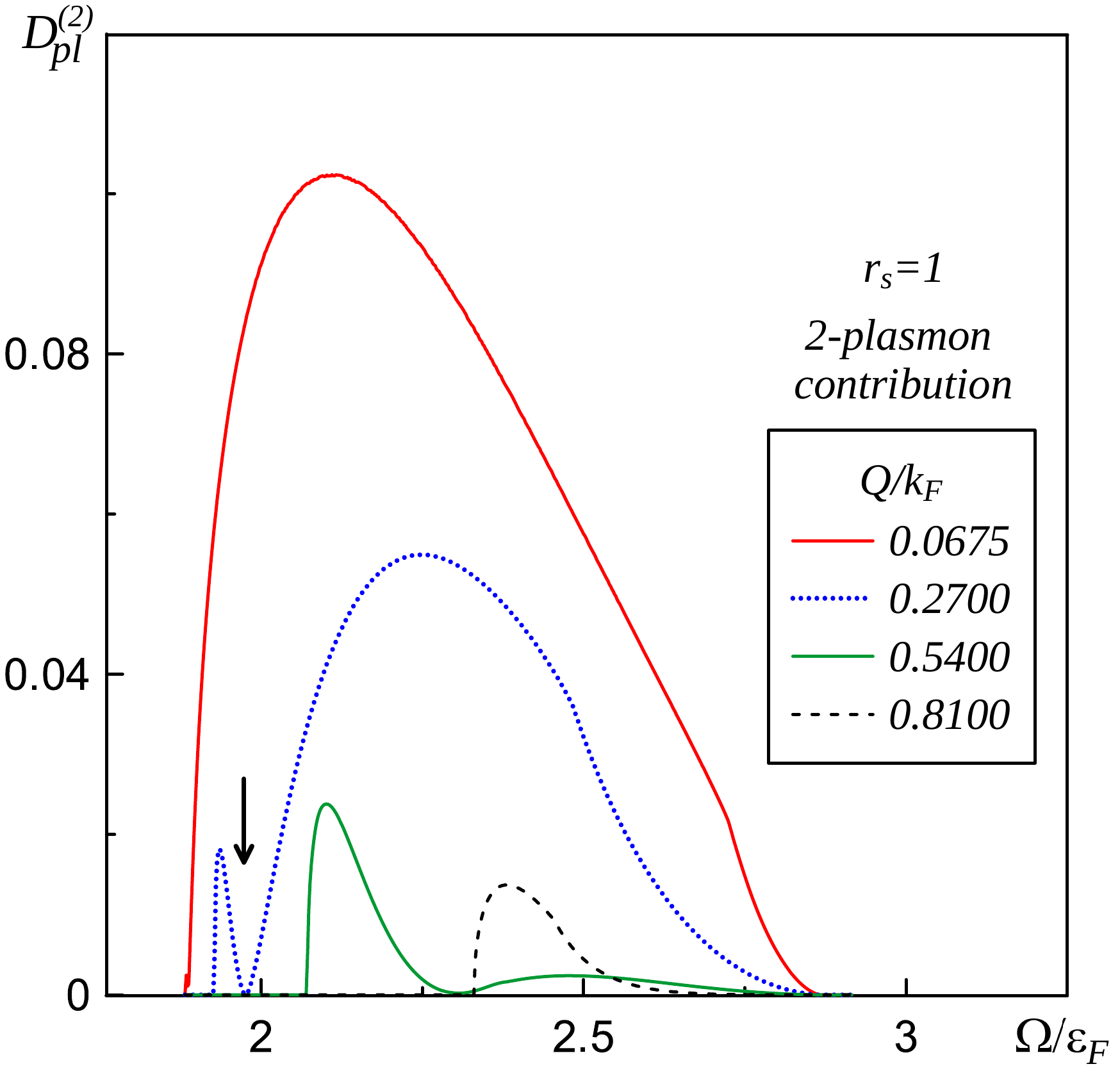}}
\subfigure{\includegraphics[scale=0.24]{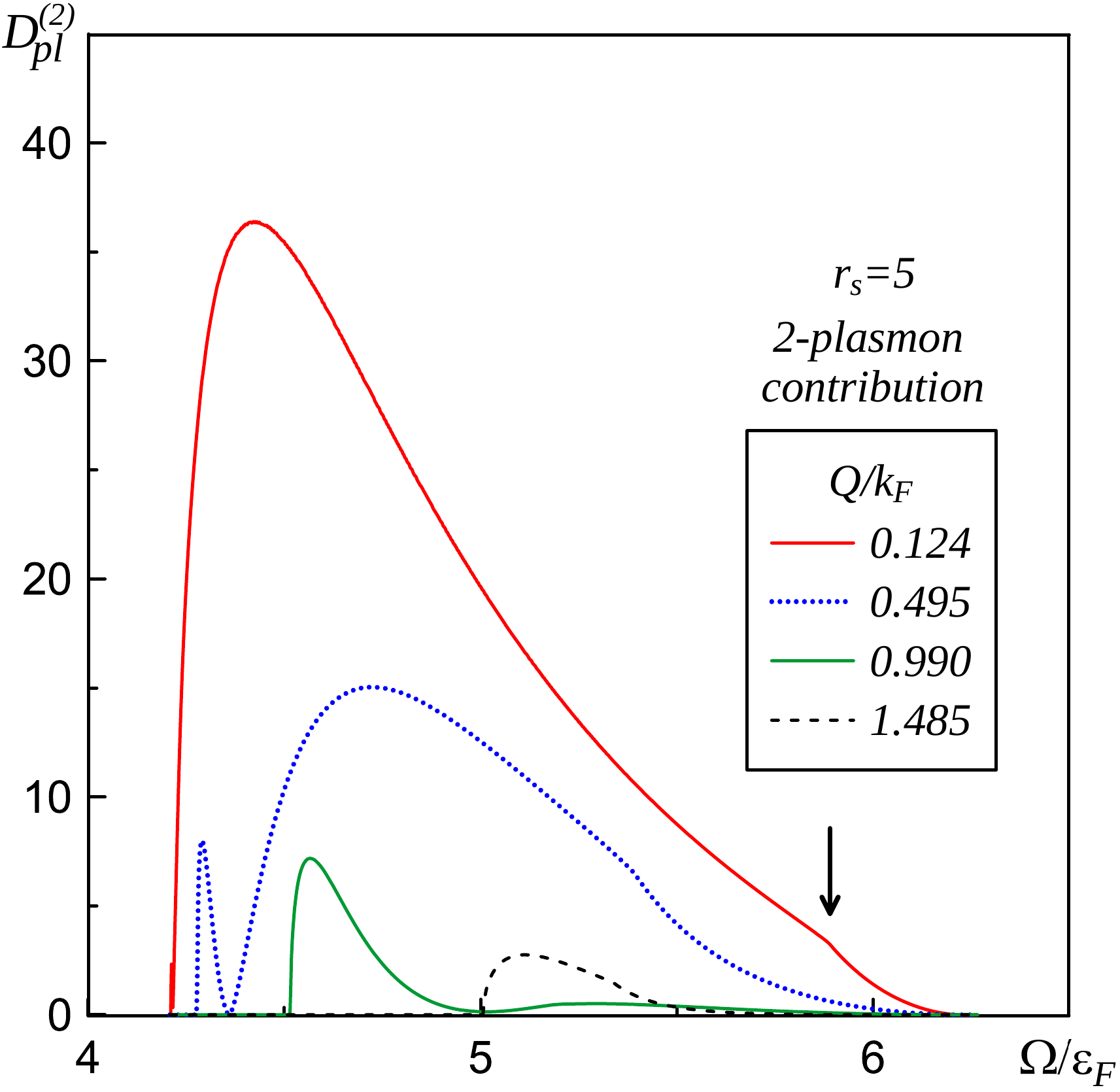}}
\caption{(color online) Two-plasmon contributions to intensity for several values of $Q$.
Left panel: $r_s=1$, the arrow points at the low-frequency minimum. Right panel: $r_s=5$, the arrow points at the high-frequency kink. }
\label{Fig08}
\end{figure}

Apart from the thresholds, there are two singular points in spectrum, corresponding to two sets of momenta ${\bf q}_1$ and ${\bf q}_2$ of two plasmons: $(q_i=0, \, q_j=Q)$ and $(q_i=Q_m, \, q_j=Q_m-Q$), where $i \ne j =1,2$. At the first singular point the spectrum vanishes because $f_2({\bf q}_1,{\bf q}_2)=0$, see Eqs.~(\ref{D2})-(\ref{f2}). This minimum is located at
\be
\Omega_{lm} = \Omega_{pl} + \omega_{pl}(Q)\, .
\label{OmegaLM}
\ee
and is clearly seen on the two-plasmon curves in their lower-frequency part, see both panels of Fig.~\ref{Fig08}, and is preceded by a maximum for obvious reasons. When $Q \to 0$, the minimum is approaching the low-energy threshold and the spectral weight displays a sharp low-amplitude oscillation. At $Q > Q_m/2$ the minimum broadens and at $Q > Q_m$ completely reshapes the spectrum.

At the second singular point there is no special reason for the two-plasmon spectrum to vanish; instead, it is seen as a kink located at
\be
\Omega_{rk} = \Omega_{m} + \omega_{pl}(Q_m-Q) \, .
\label{OmegaRK}
\ee
It is more pronounced on small-to-moderate-$Q$ curves (when the low-frequency maximum is smaller in intensity than the central one) in their high-frequency part, see both panels of Fig.~\ref{Fig08}. At $Q \to 0$ the kink is approaching the high-frequency threshold at $2\Omega_m$, while for $Q > Q_m$ it disappears together with the central maximum.

\begin{figure}[htp]
\centering
\subfigure{\includegraphics[scale=0.24]{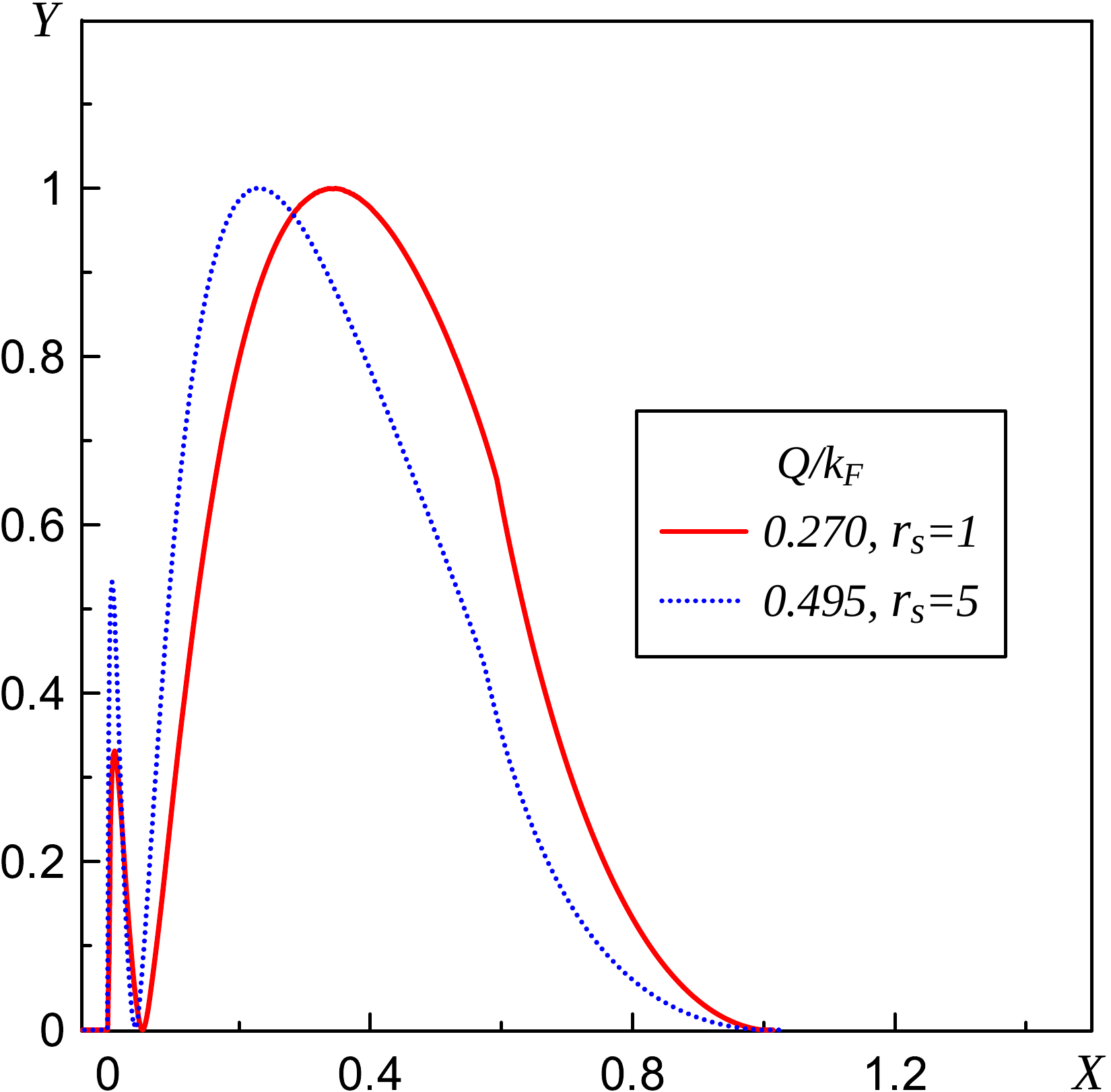}}
\subfigure{\includegraphics[scale=0.24]{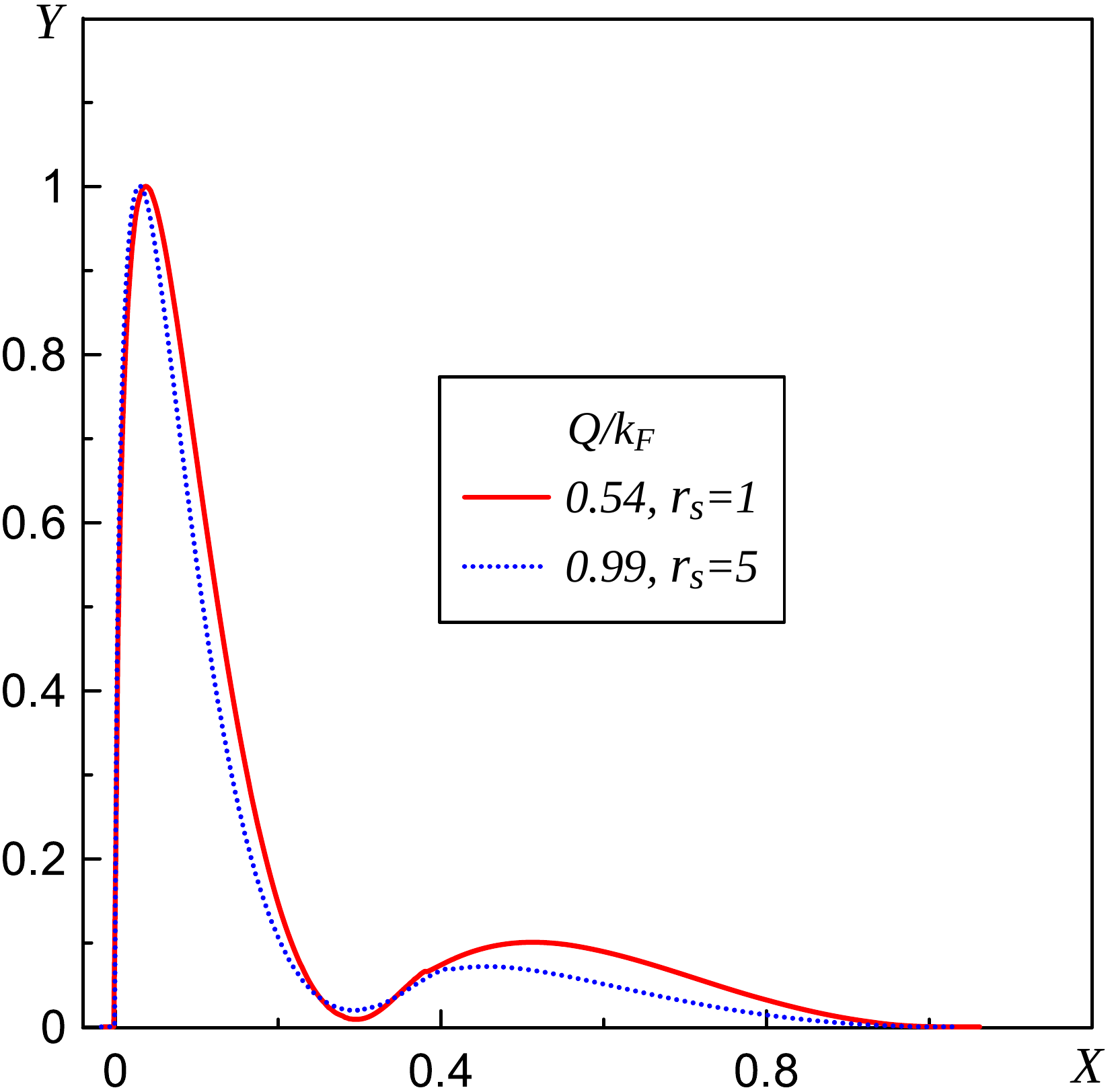}}
\caption{(color online) Scaled two-plasmon intensities $Y=D^{(2)}_{pl}/\{ D^{(2)}_{pl} \}_{max}$ as functions of $X=[\Omega-2\Omega(Q/2)]/2[\Omega_m-\Omega(Q/2)] \in [0,1]$ for $r_s = 1$ and $r_s=5$.
The momentum transfer for two cases was adjusted to have the same ratio $Q/Q_m(r_s)$. }
\label{Fig09}
\end{figure}

Disregarding the overall amplitude of the signal, the two-plasmon curves for $r_s=1$ and $5$ look very similar and in RPA the non-perturbative values of $r_s$ do not introduce new features to the two-plasmon spectra. To quantify this statement further, in Fig.~\ref{Fig09} we compare spectra for $r_s=1$ and $r_s=5$ using properly scaled variables: the momentum transfer was chosen to have the same ratio for $Q/Q_m(r_s)$, the intensity was normalized to unity at the maximum, and the frequency was scaled to be in the $[0,1]$ interval.

It is also worth mentioning that for the two-plasmon process, the intensity at the central maximum saturates to a constant when $Q \to 0$ while the spectrum remains broad, in contrast to the $\propto Q^2$ spectral weight scaling associated with the single-plasmon process.

%%%%%%%%%%%%%%%%%%%%%%%%%%%%%%%%%%%%%%%%%%%%%%%%%%%%%%%%%%%%%%%%%%%%%%%%%%%%%%%%%%%%%%%%%%%%%%%%%
\subsection{Hybrid processes}
\label{HYB}
%%%%%%%%%%%%%%%%%%%%%%%%%%%%%%%%%%%%%%%%%%%%%%%%%%%%%%%%%%%%%%%%%%%%%%%%%%%%%%%%%%%%%%%%%%%%%%%%%

\begin{figure}[tbh]
\centering
\subfigure{\includegraphics[scale=0.24]{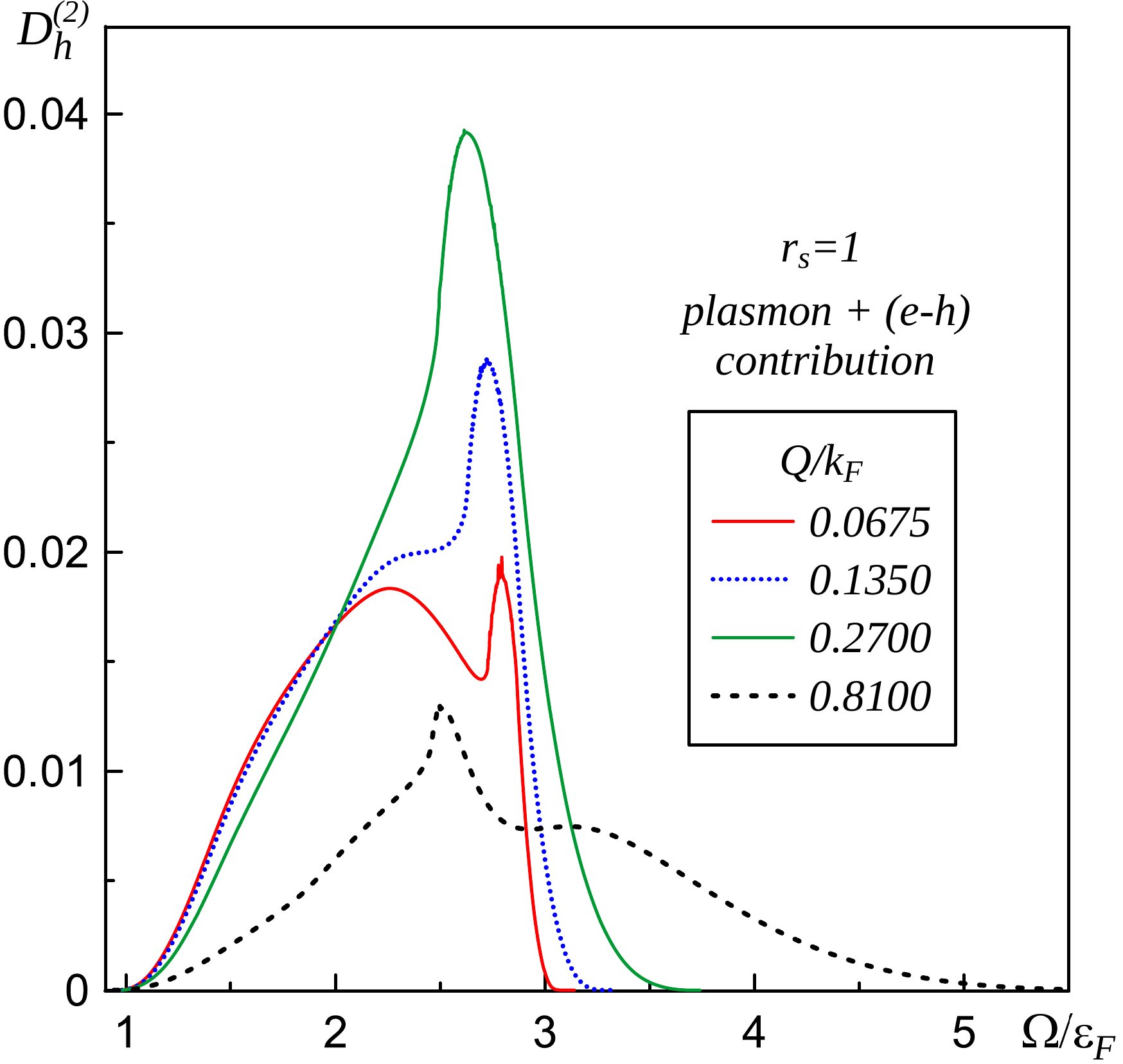}}
\subfigure{\includegraphics[scale=0.24]{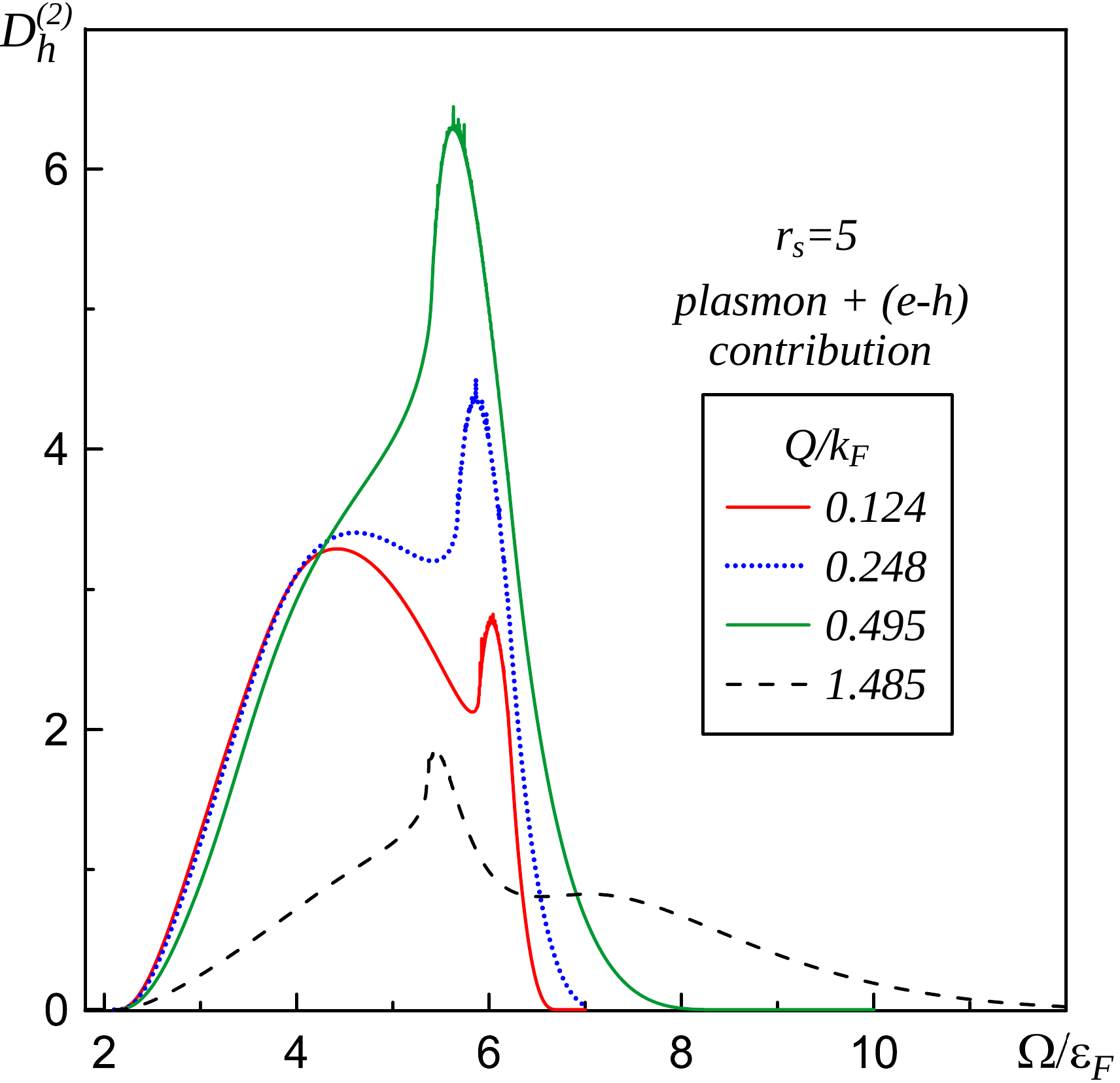}}
\caption{(color online) Hybrid contribution to intensity as a function of $\Omega$
for several values of $Q$ at $r_s=1$ (left panel) and $r_s=5$ (right panel). }
\label{Fig10}
\end{figure}

The thresholds for the hybrid spectrum are at $\Omega_{pl}$ and $\Omega_m + \Omega_{e-h}(Q_m+Q)$, as dictated by the energy-momentum conservation laws and limited momentum support for the plasmon dispersion. At $Q \to 0$ the upper threshold is located at $2\Omega_m$ (see Eq.~\ref{WEH2}). In Fig.~\ref{Fig10} we show the hybrid spectra for $r_s=1$ and $r_s=5$. Since the lower threshold is at the plasma frequency $\Omega_{pl}$, the hybrid spectrum overlaps either with the single-plasmon, or with the single-pair spectrum. At $Q \to 0$ the intensity of the broad central maximum saturates to a constant, in contrast to the $\propto Q^3$ scaling of the shrinking single-pair spectrum and $\propto Q^2$ scaling of the single-plasmon weight. However, since the hybrid intensity vanishes at $\Omega \to \Omega_{pl}$, it does not obscure the sharp single-plasmon peak in the $Q \to 0$ limit (contrary to the two-pair process discussed next).

\begin{figure}[tbh]
\centerline{\includegraphics[angle = 0,width=0.8\columnwidth]{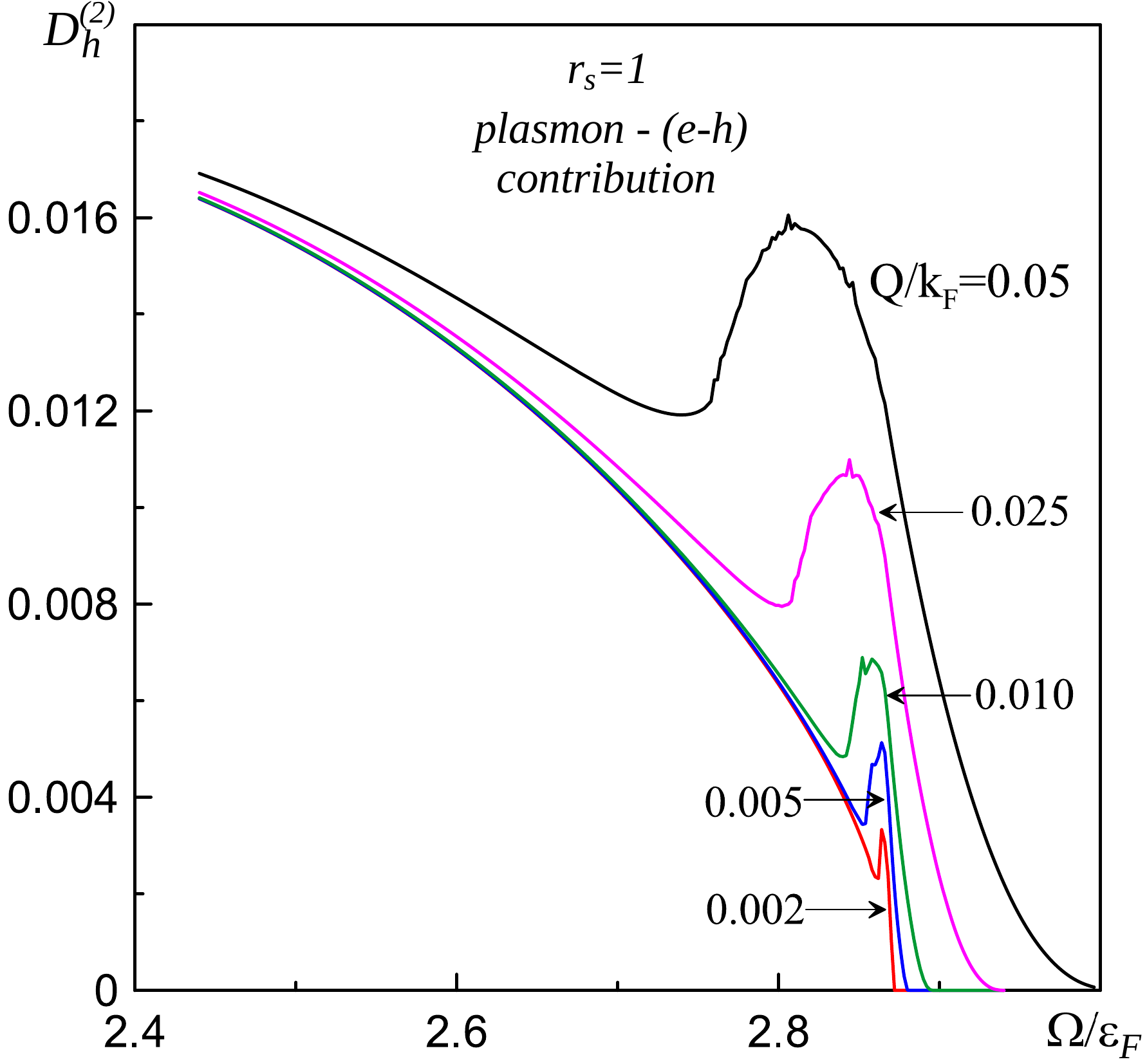}}
\caption{(color online) Hybrid contribution to intensity: $D^{(2)}_{h}$ as a function of $\Omega$ at small values of $Q$ and $r_s=1$.}
\label{Fig11}
\end{figure}

At $Q > 0$ the high-frequency peak emerges from the threshold, grows in amplitude, see Fig.~\ref{Fig11}, and ultimately reshapes the entire curve, see Fig.~\ref{Fig10}. The origin of this peak at small $Q$ can be traced to the fact that near the upper threshold the pair and plasmon excitations both have momenta close to $|Q_m|$, but pointing in the opposite directions. The peak develops from the interplay between the singularities in the plasmon peak residue and pair intensity when $\Omega \to \Omega_m$, see Figs.~\ref{Fig04} and \ref{Fig05}, and the available phase-space volume. It is not an accident that the emerging hybrid peak location correlates with the location of the kink on the two-plasmon curve because for $Q\to Q_m$ the single-pair spectrum intensity diverges on approach to $\Omega_m$ mimicking a plasmon resonance.

\begin{figure}[tbh]
\subfigure{\includegraphics[scale=0.24]{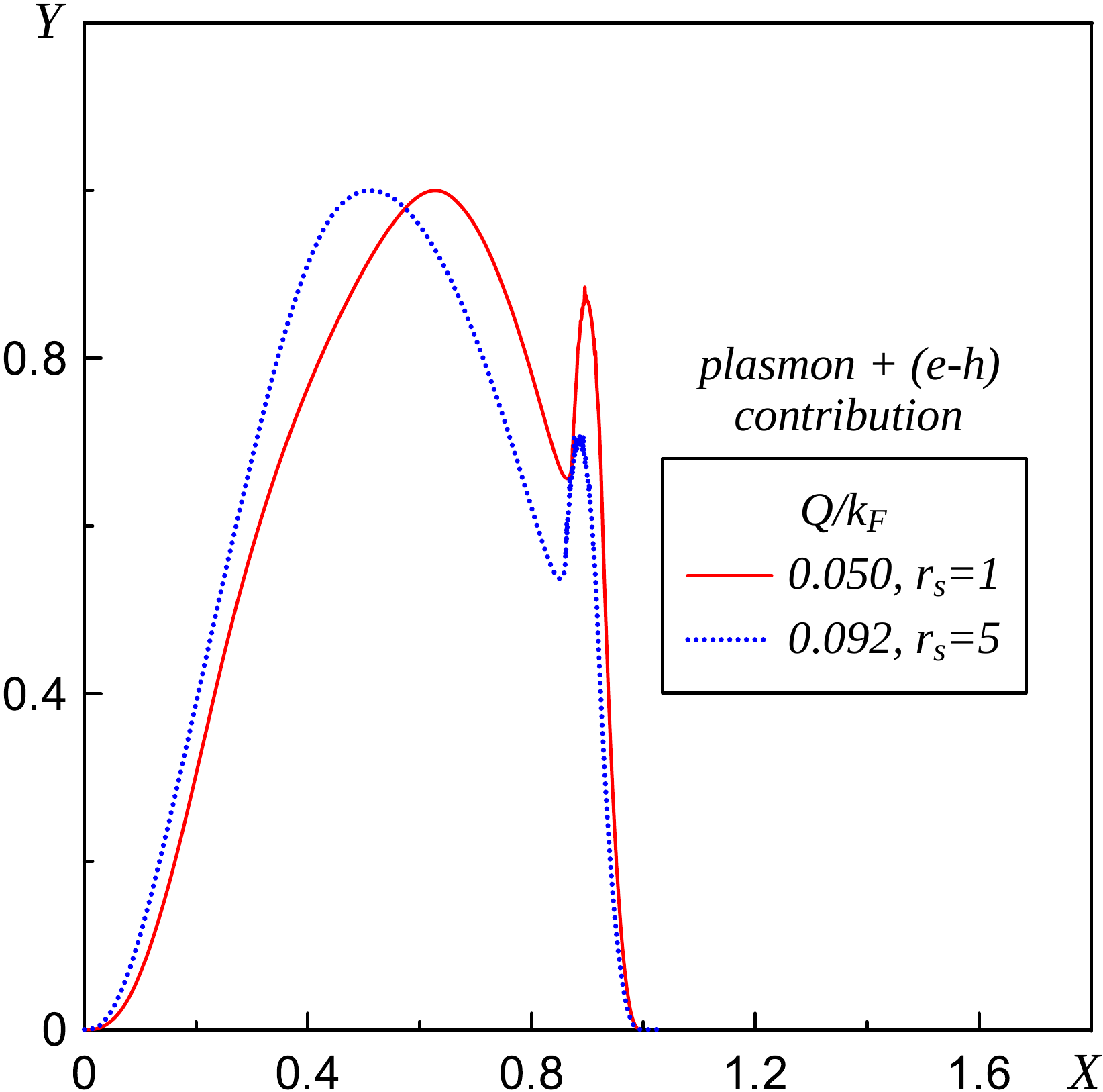}}
\subfigure{\includegraphics[scale=0.24]{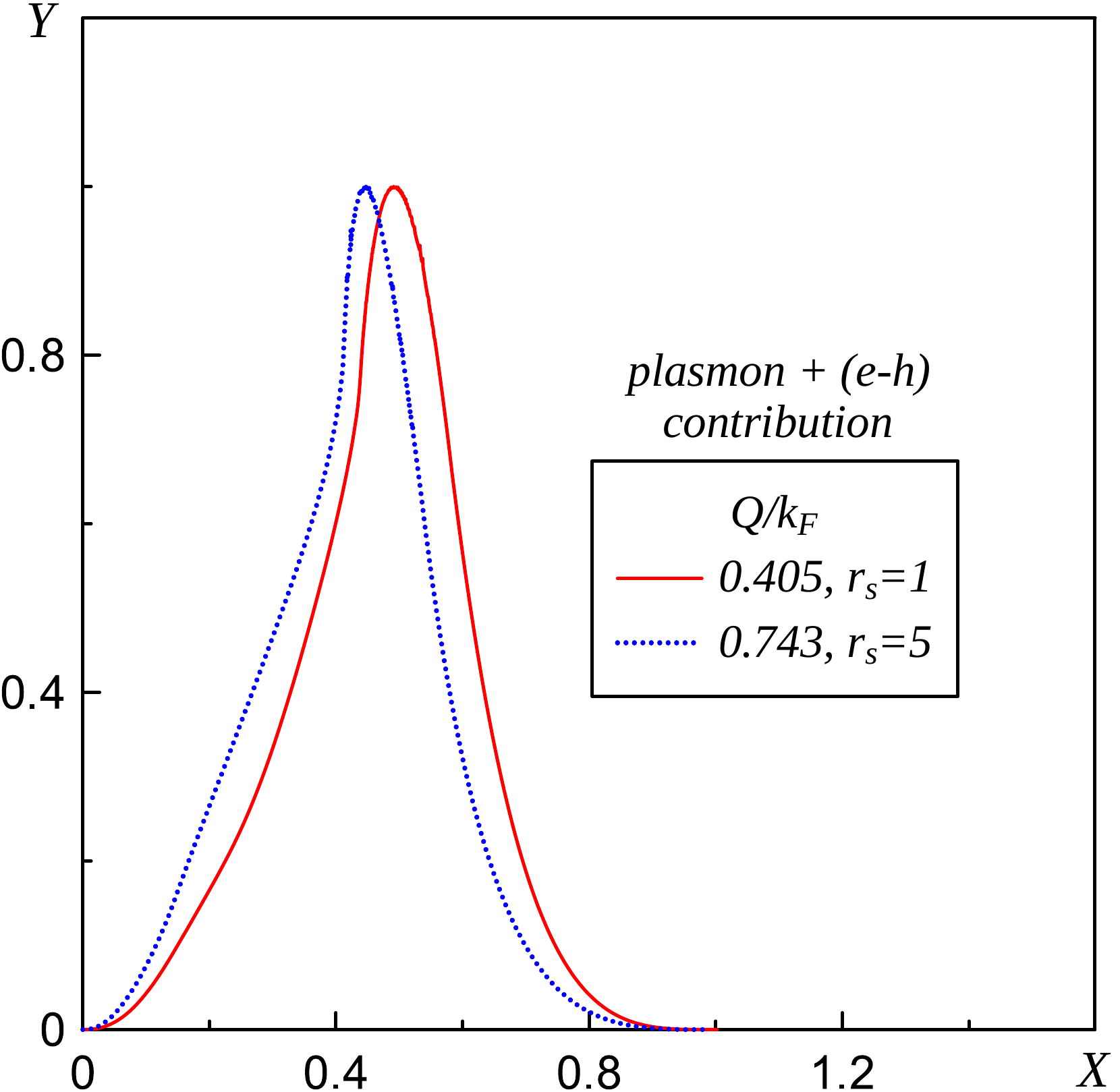}}
\caption{(color online) Scaled hybrid contributions to intensity for $r_s=1$ and $r_s=5$. The data
are plotted for $Y=D^{(2)}_{h}/\{ D^{(2)}_{h} \}_{max}$ as functions of
$X=[\Omega-\Omega_{pl}]/[\Omega_m + \Omega_{e-h}(Q_m+Q)-\Omega_{pl}] \in [0,1]$.
The momentum transfer for two cases was adjusted to have the same ratio $Q/Q_m$. }
\label{Fig12}
\end{figure}

Again, the curves for different values of $r_s$ in Fig.~\ref{Fig10} appear similar, and this observation can be quantified by plotting the data using scaled variables as it was done for the two-plasmon process. The result is presented in Fig.~\ref{Fig12}. The intensity of the hybrid process does not vanish at $Q \to 0$; however, its features are difficult to observe because of interplay with the two-pair process discussed below. Nevertheless, the upper threshold of the hybrid contribution is visible on total curves at small values of momentum transfer which gives access to information about the single $e-h$ process.

%%%%%%%%%%%%%%%%%%%%%%%%%%%%%%%%%%%%%%%%%%%%%%%%%%%%%%%%%%%%%%%%%%%%%%%%%%%%%%%%%%%%%%%%%%%%%%%%%
\subsection{Two-pair processes}
\label{2EH}
%%%%%%%%%%%%%%%%%%%%%%%%%%%%%%%%%%%%%%%%%%%%%%%%%%%%%%%%%%%%%%%%%%%%%%%%%%%%%%%%%%%%%%%%%%%%%%%%%

\begin{figure}[tbh]
\subfigure{\includegraphics[scale=0.24]{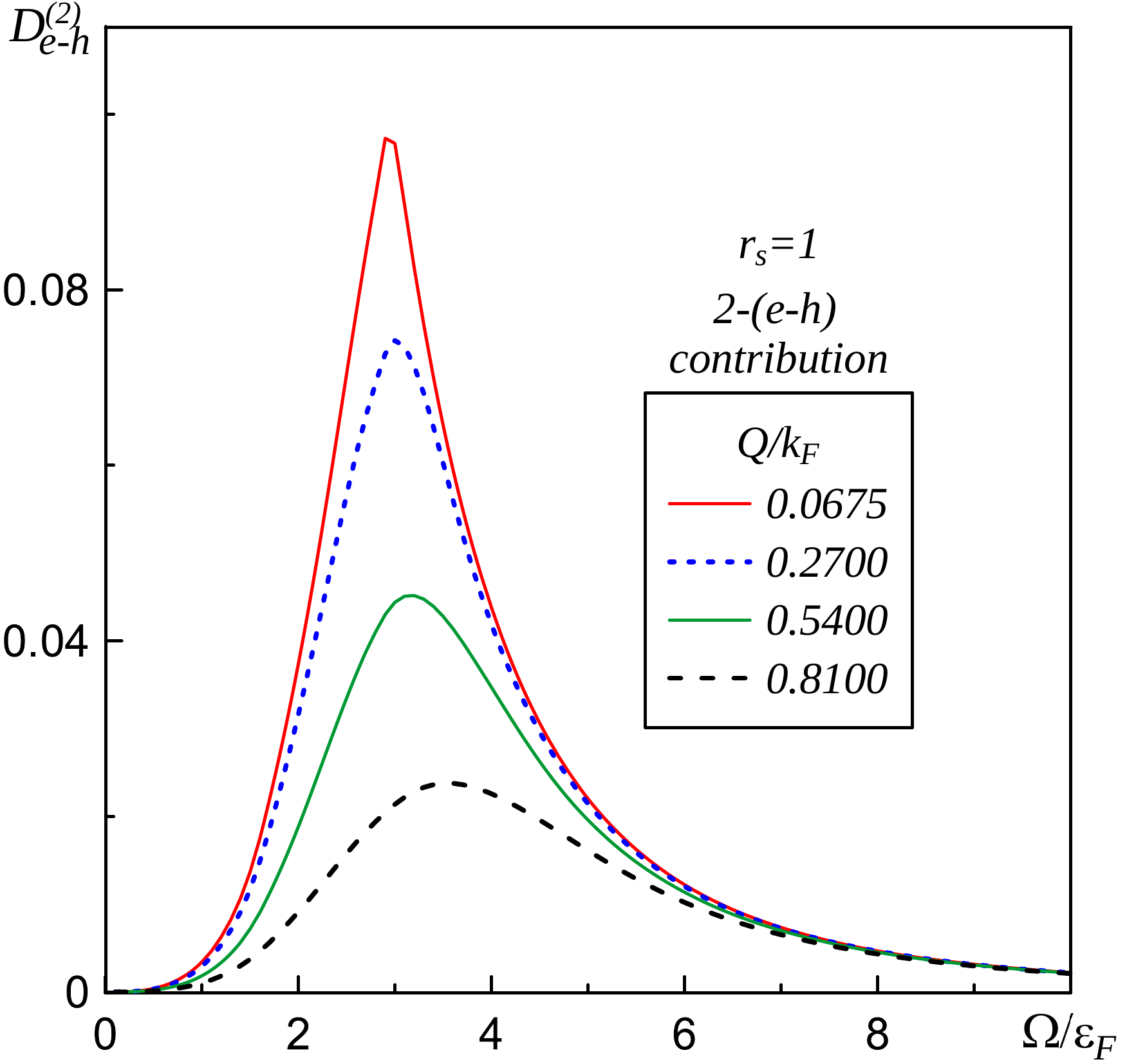}}
\subfigure{\includegraphics[scale=0.24]{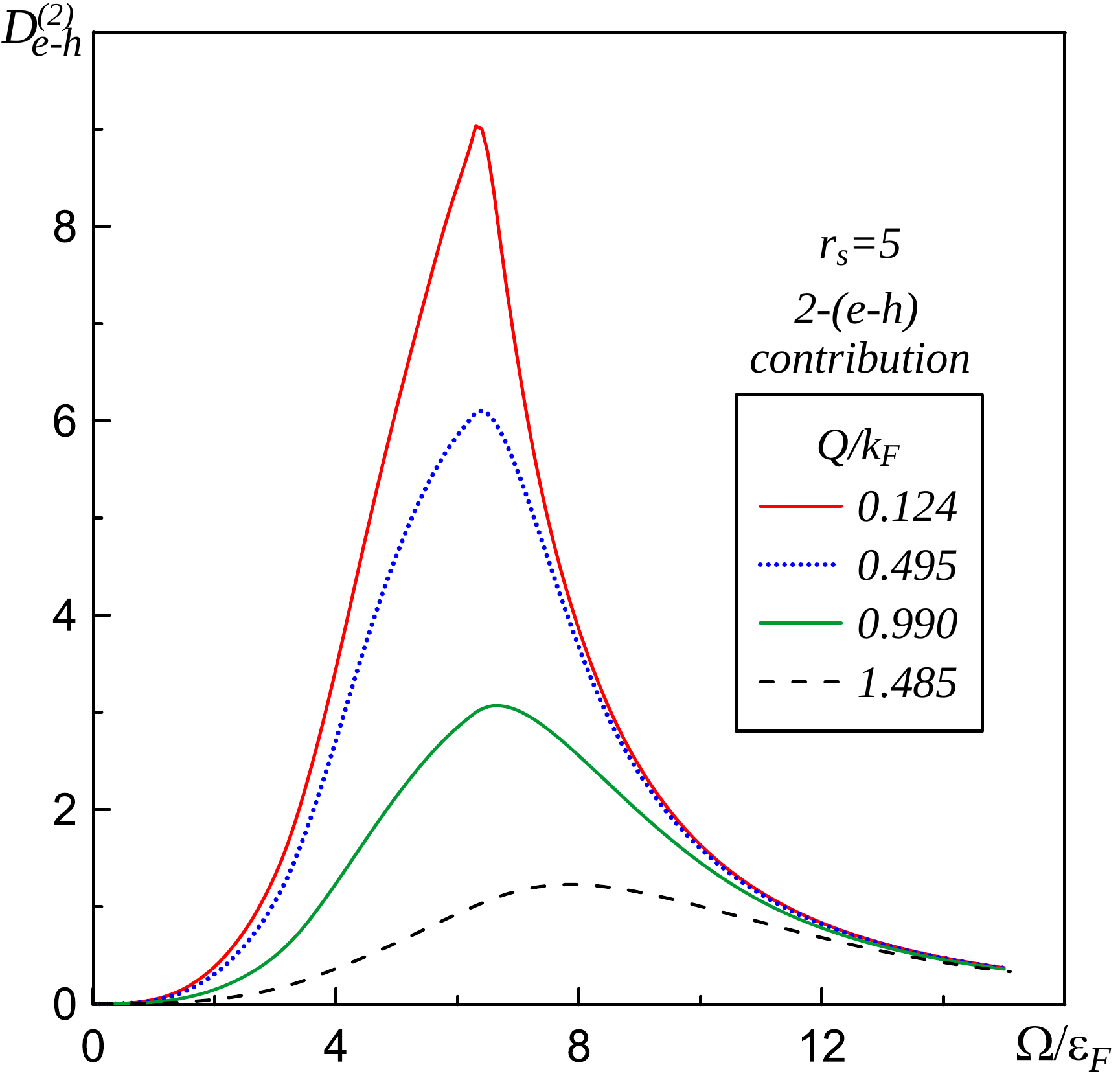}}
\caption{(color online) Two-pair contribution intensity as a function of $\Omega$
for several values of $Q$ at $r_s=1$ (left panel) and $5$ (right panel). }
\label{Fig13}
\end{figure}

The last process contributing to the second-order spectra is the two-pair one. The result is expected to be a smooth peak. The typical shapes are presented in Fig.~\ref{Fig13}---they start at $\Omega=0$ and at large frequency demonstrate an asymptotic $\sim \Omega^{-7/2}$ decay. At small frequencies, the signal is proportional to $\Omega^2$, as expected from the single-pair intensity $\propto \Omega$ at $\Omega \to 0$.

The maximum at small momenta is located close to $2\Omega_m$, where the two-plasmon and hybrid processes have the kink and the high-frequency peak, respectively. The intensity maximum saturates to a constant in the $Q \to 0$ limit, implying that in this limit the entire RIXS spectrum is dominated by the second-order processes. By comparing the two-pair contribution to other second-order processes, we see that the former dominates the total two-excitation spectrum in the low- and high-frequency parts. By comparing curves in the left and right panels of Fig.~\ref{Fig13} we observe similar universality as for other second-order processes.

%%%%%%%%%%%%%%%%%%%%%%%%%%%%%%%%%%%%%%%%%%%%%%%%%%%%%%%%%%%%%%%%%%%%%%%%%%%%%%%%%%%%%%%%%%%%%%%%%%%
\subsection{Total two-excitation intensities}
%%%%%%%%%%%%%%%%%%%%%%%%%%%%%%%%%%%%%%%%%%%%%%%%%%%%%%%%%%%%%%%%%%%%%%%%%%%%%%%%%%%%%%%%%%%%%%%%%%%

\begin{figure}[tbh]
\subfigure{\includegraphics[scale=0.24]{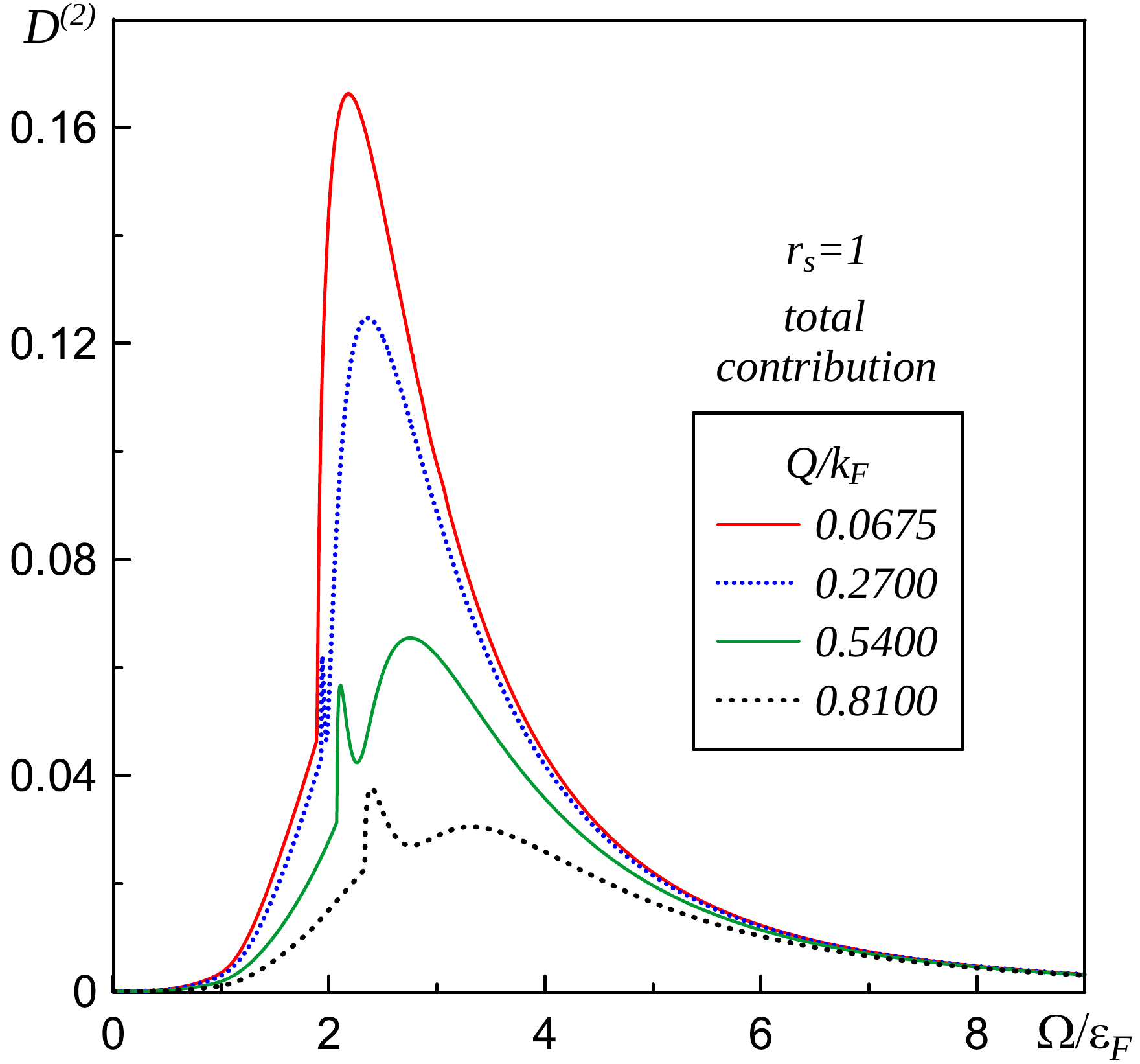}}
\subfigure{\includegraphics[scale=0.24]{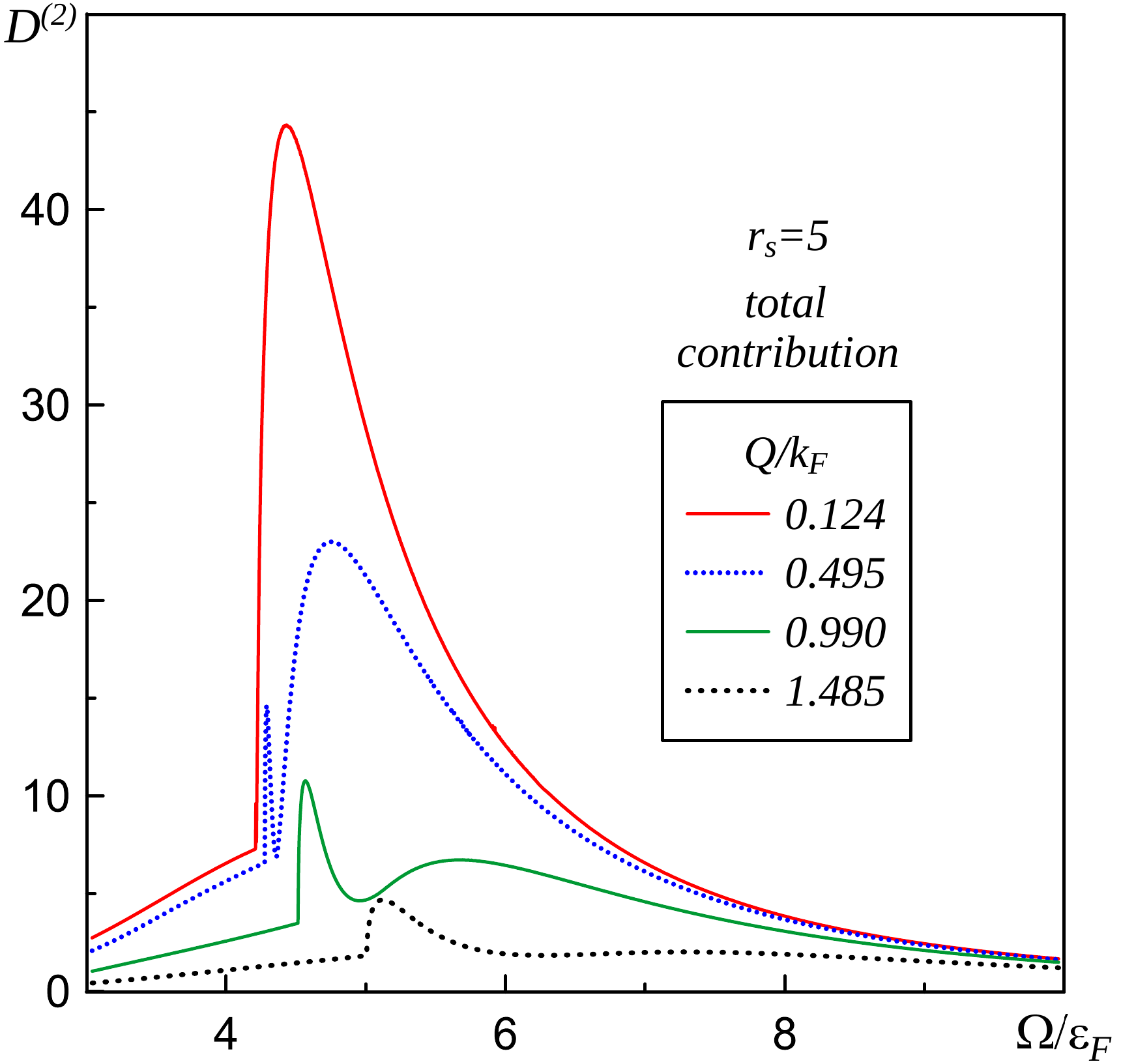}}
\caption{(color online) Total second-order intensity as a function of $\Omega$
for several values of $Q$ at $r_s=1$ (left panel) and $5$ (right panel). }
\label{Fig14}
\end{figure}

\begin{figure}[tbh]
\subfigure{\includegraphics[scale=0.24]{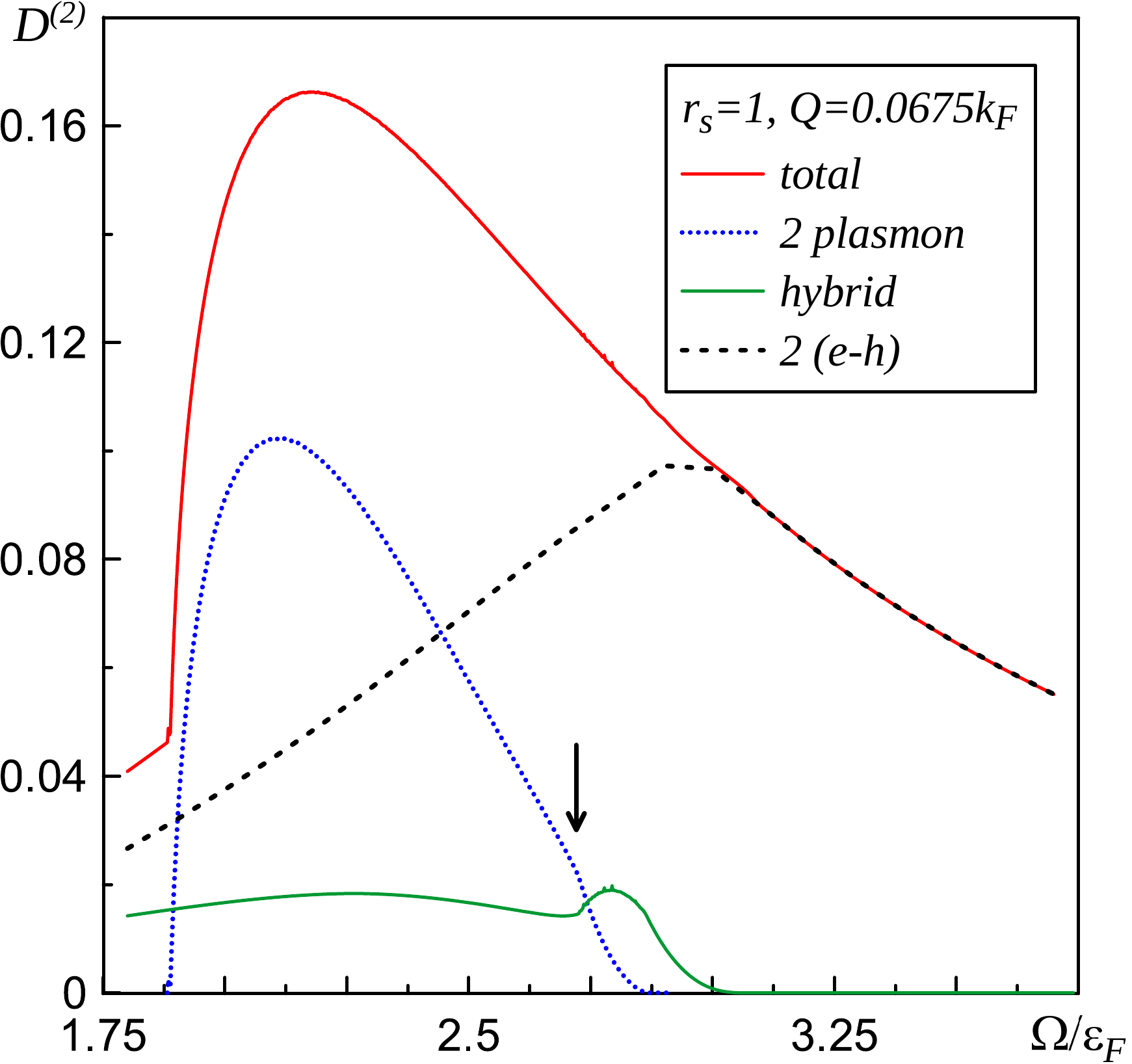}}
\subfigure{\includegraphics[scale=0.24]{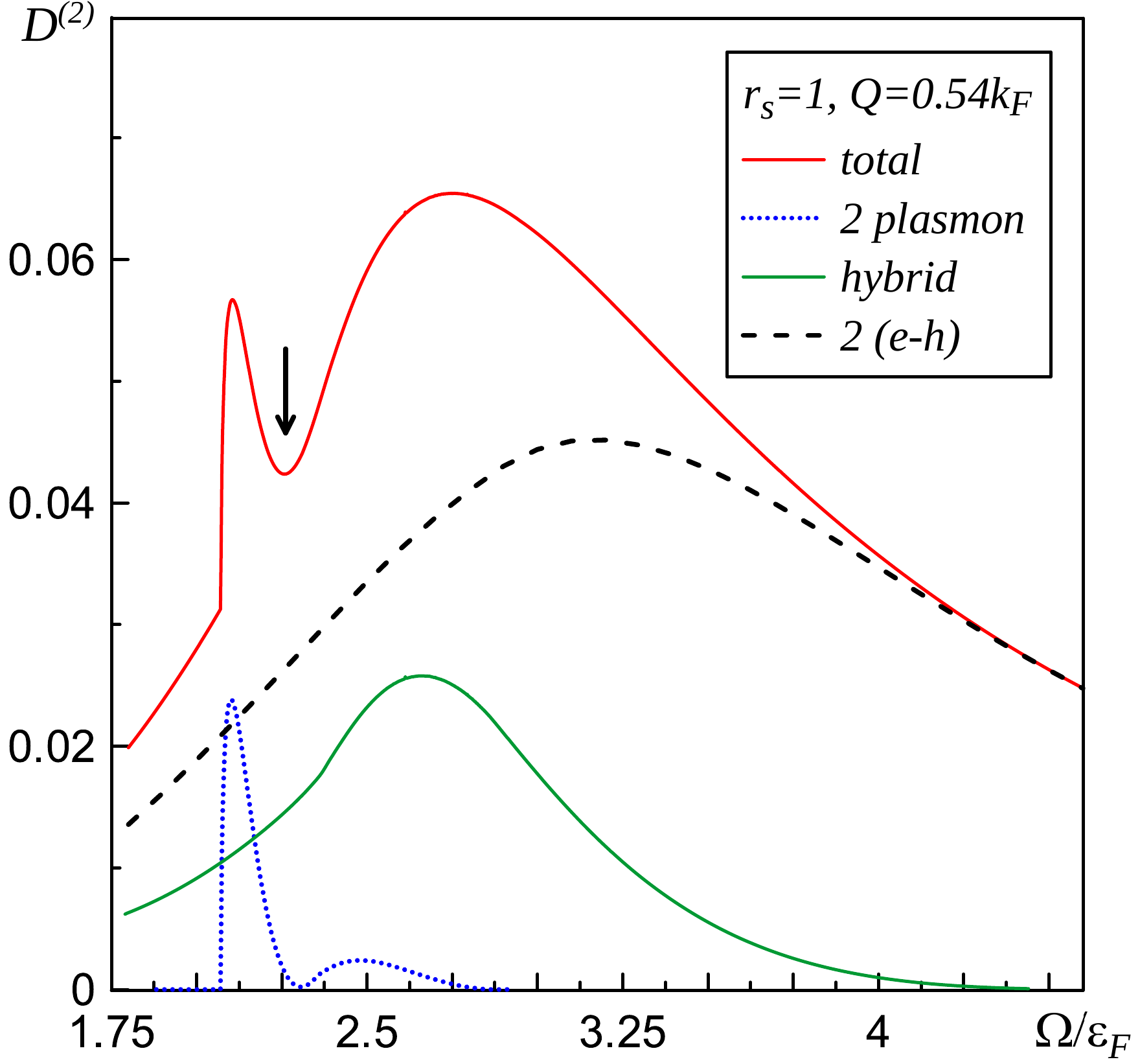}}
\subfigure{\includegraphics[scale=0.24]{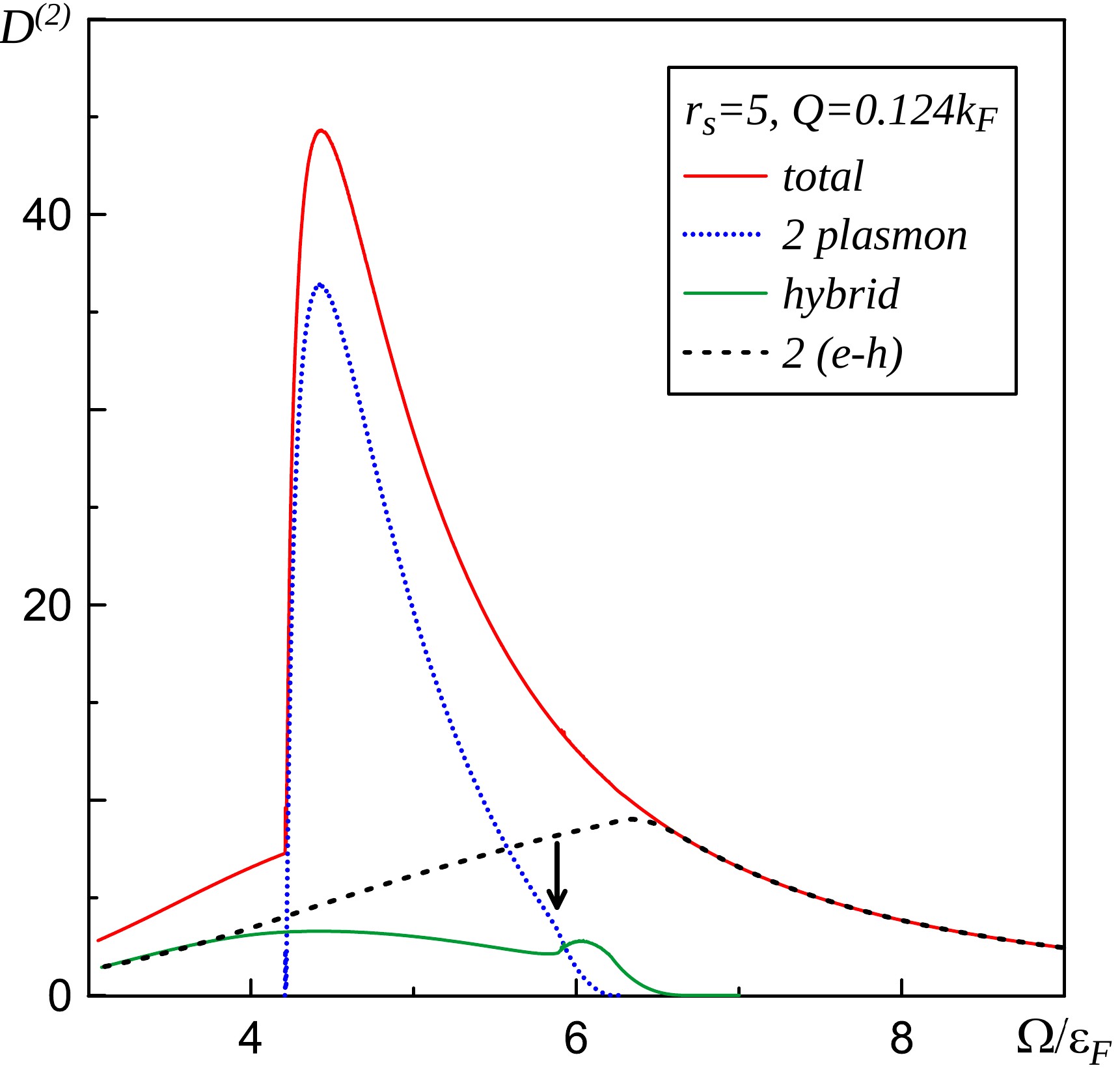}}
\subfigure{\includegraphics[scale=0.24]{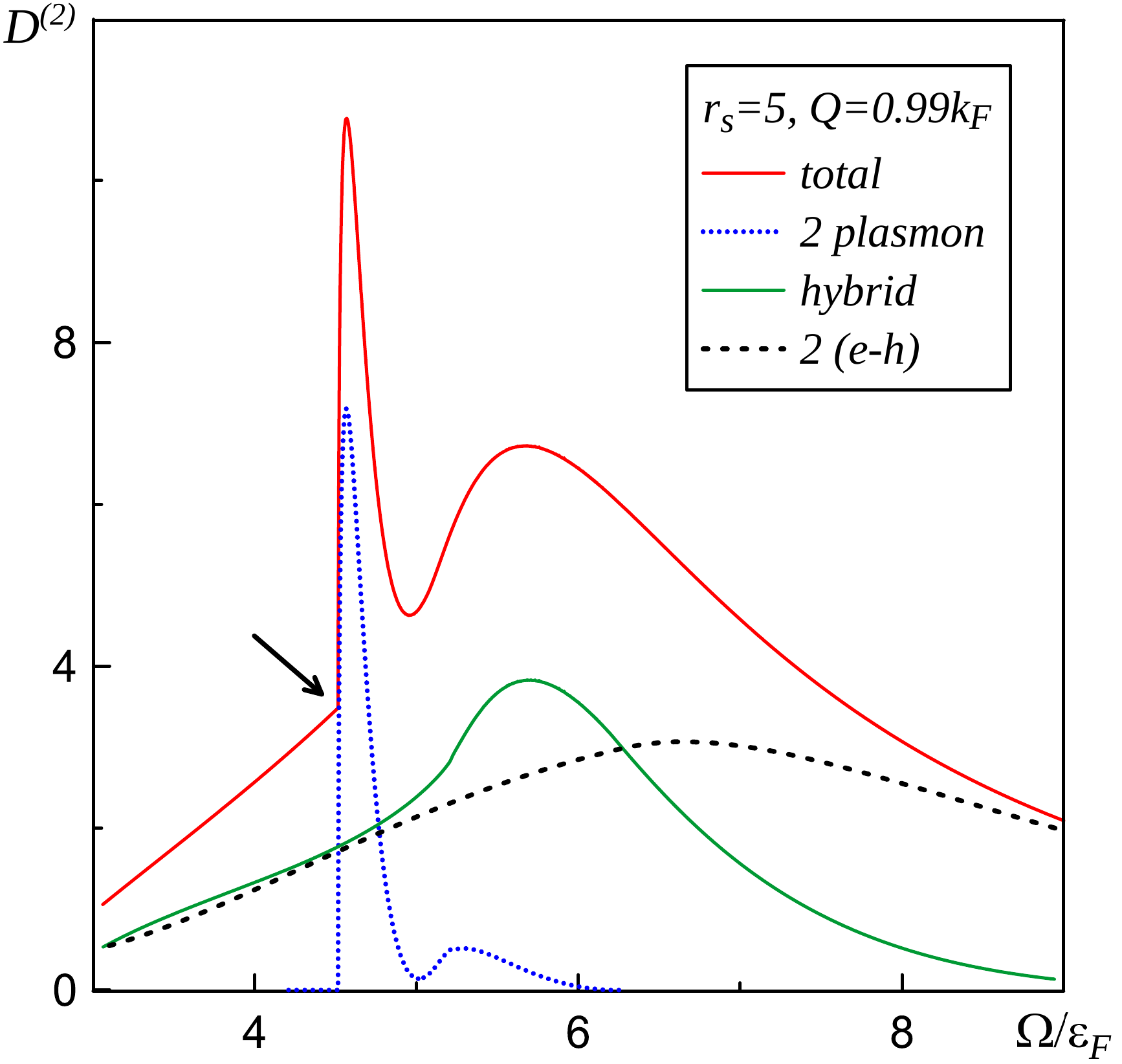}}
\caption{(color online) Total and all partial (two-plasmon, two-pairs, and hybrid) contributions to intensity as a function of $\Omega$. Upper left panel: $r_s=1$ and $Q=0.0675 k_F$; the arrow points at the high-frequency kink on the two-plasmon curve. Upper right panel: $r_s=1$ and $Q=0.54 k_F$; the arrow points at the low-frequency minimum. Lower left panel: $r_s=5$ and $Q=0.124 k_F$;
the arrow points at the high-frequency kink on the two-plasmon curve. Lower right panel: $r_s=5$ and $Q=0.99 k_F$; the arrow points at the low-frequency kink. }
\label{Fig15}
\end{figure}

By combining all second-order processes we obtain the total two-excitation intensity, see Fig.~\ref{Fig14}. It is expected that some of the features clearly seen on individual curves may be masked when different contributions overlap. At low ($\Omega \lesssim \Omega_{pl}$) and high ($\Omega \gtrsim 2\Omega_m$) frequencies the total signal is dominated by the two-pair process. At intermediate frequencies the leading contribution often comes from the two-plasmon process which is responsible for sharp features at $Q \lesssim Q_m$, see Figs. \ref{Fig14} and \ref{Fig15}.

In Fig.~\ref{Fig15} we present the central part of the spectrum and compare all partial and total signals side-by-side in order to understand how spectral features in the final result should be ``decoded". The low-frequency kink and minimum are due to the two-plasmon process and these sharp features remain clearly visible, while the lower threshold for the hybrid process is masked by the two-pair contribution. The high frequency kink in the two-plasmon spectrum (best seen in two left panels in Fig.~\ref{Fig15}) is compensated by the non-monotonous dependence of the hybrid process and is not visible on total curves.

\begin{figure}[tbh]
\subfigure{\includegraphics[scale=0.24]{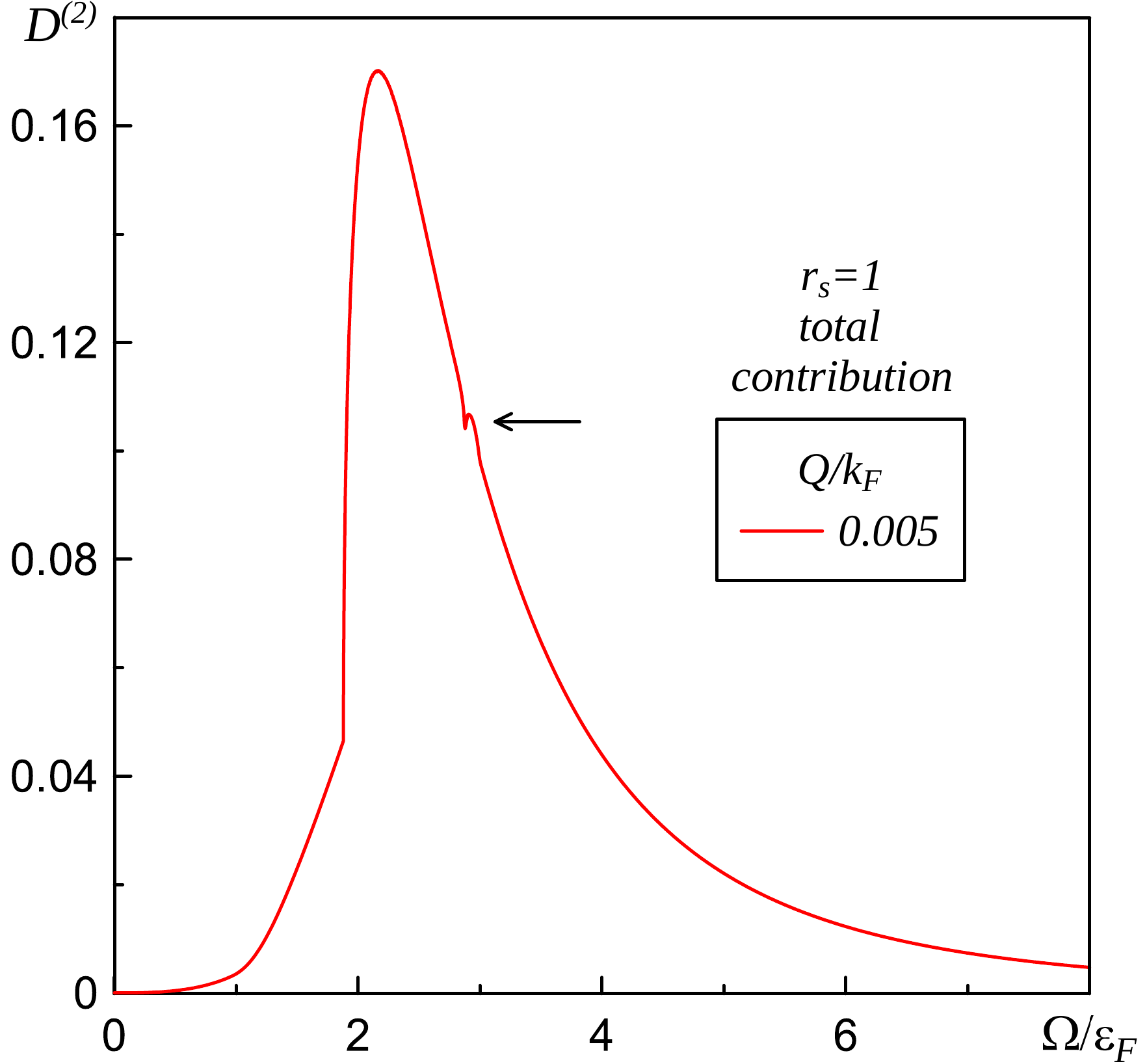}}
\subfigure{\includegraphics[scale=0.24]{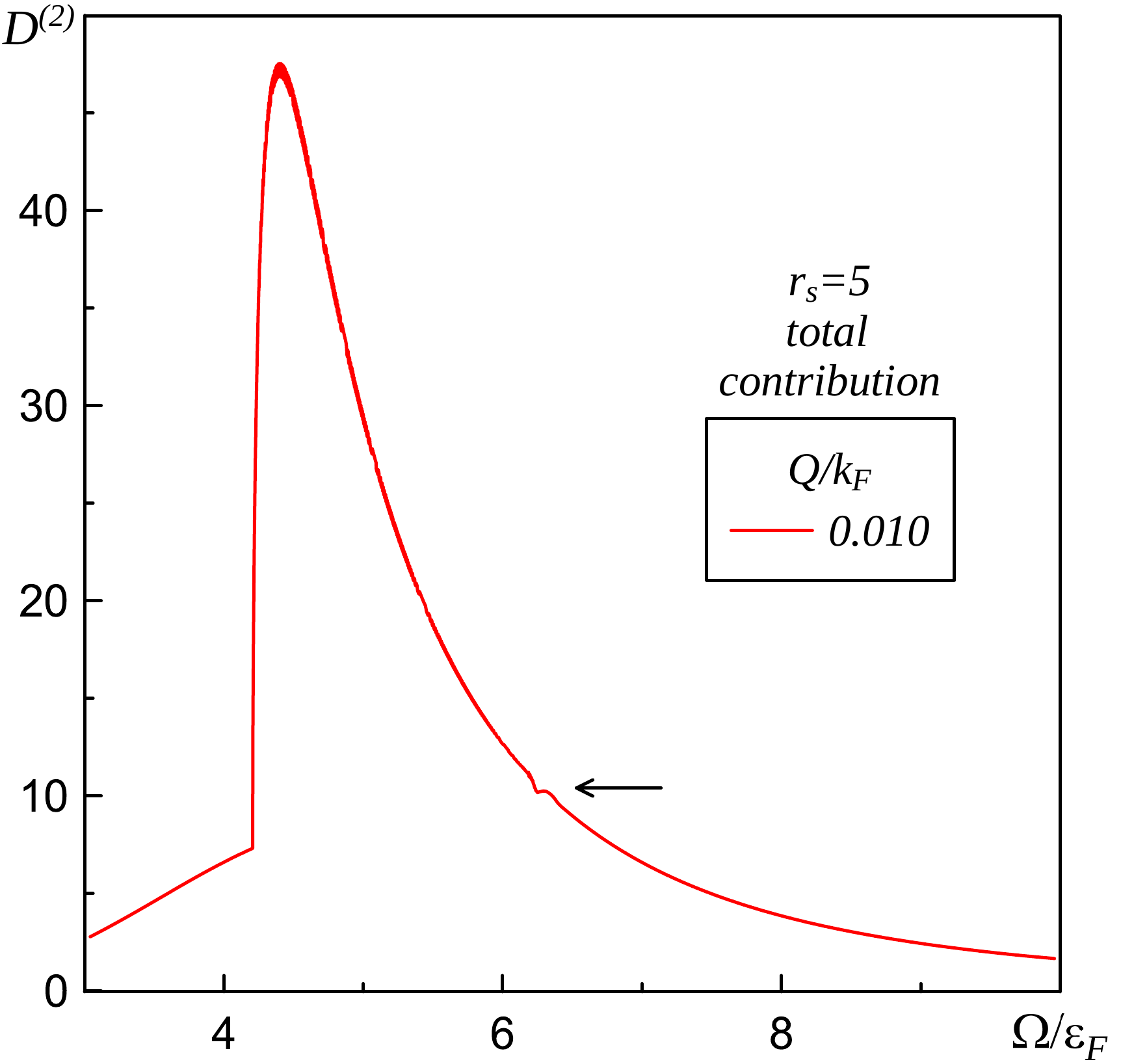}}
\caption{ (color online) Total second-order intensity as a function of $\Omega$
for small values of $Q$ at $r_s=1$ (left panel) and $5$ (right panel). The arrows point at the wiggle-like spectral anomaly appearing at small momentum transfer as the result of interplay between the high frequency hybrid signal near its threshold and the two-pair intensity maximum.}
\label{Fig16}
\end{figure}

\begin{figure}[tbh]
\centerline{\includegraphics[angle = 0,width=0.8\columnwidth]{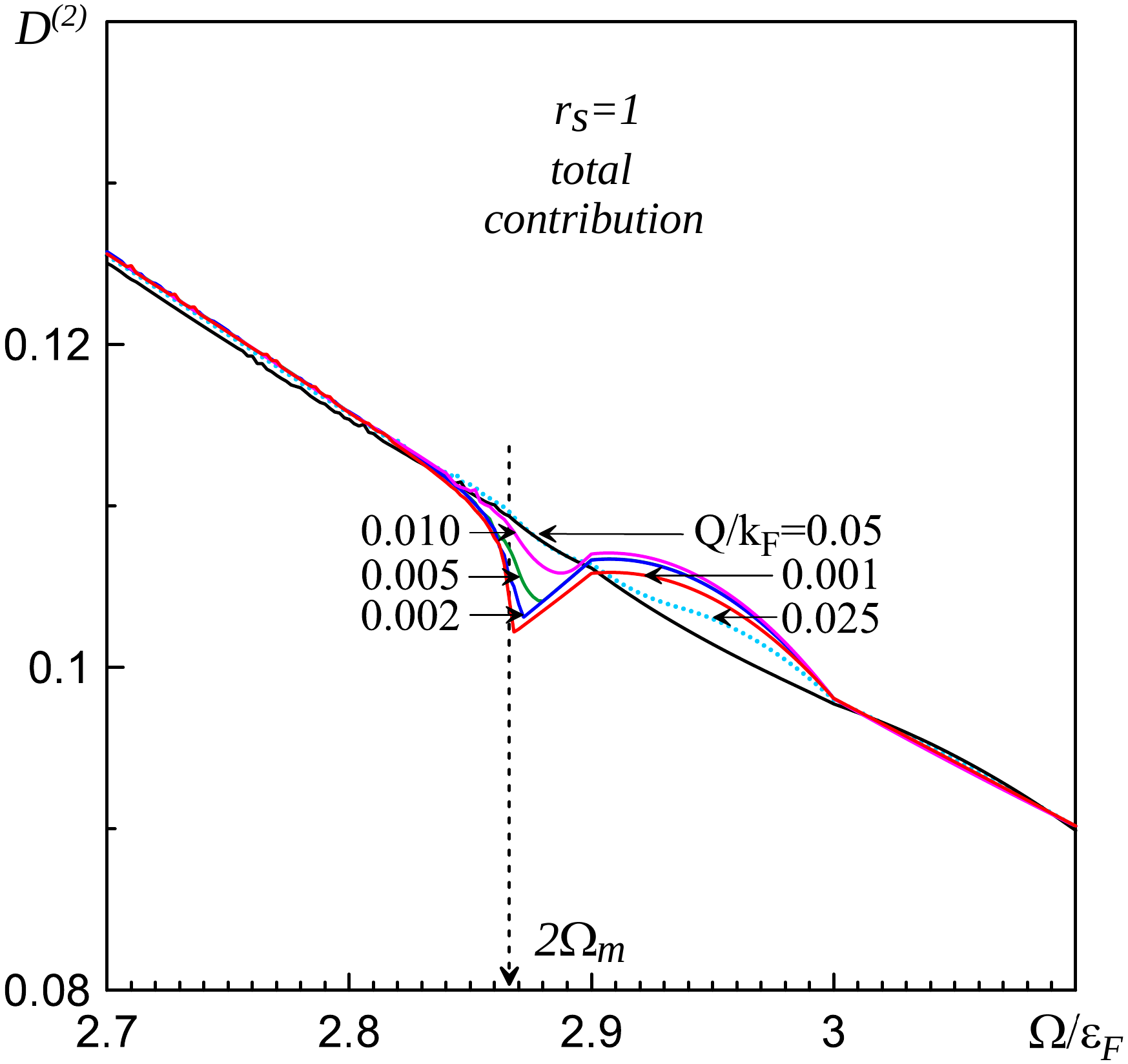}}
\caption{(color online) Total second-order intensity in the vicinity of the $2\Omega_m$ frequency for progressively smaller values of $Q$ at $r_s=1$ demonstrating the development of the wiggle-like spectral anomaly.}
\label{Fig17}
\end{figure}

The interplay between the sharp non-monotonous hybrid signal at high frequencies (see Fig.~\ref{Fig11}) and the two-pair intensity maximum results in a wiggle-like spectral anomaly seen in Fig.~\ref{Fig16} at around $2\Omega_m$. This anomaly is pronounced
only at small momenta and disappears at larger values of $Q$, see Fig.~\ref{Fig17}. The position of the developing at $Q \to 0$ minimum corresponds to the upper threshold of the hybrid process.

As it has already been mentioned, the two-plasmon high-frequency kink is not seen on the total curve, due to the interplay between the two-plasmon and hybrid processes. This is because these processes are characterized by the kink and anti-kink features that compensate each other (since the plasmon peak at $Q=Q_m$ is getting absorbed by the $e-h$ continuum without change in the total spectral weight).

Obviously, the universality of all features established for partial contributions carries through to the total intensity. However, as it has already been mentioned, the overlap of partial contributions at different values of momenta transfer may mask some features.

%%%%%%%%%%%%%%%%%%%%%%%%%%%%%%%%%%%%%%%%%%%%%%%%%%%%%%%%%%%%%%%%%%%%%%%%%%%%%%%%%%%%%%%%%%%%%%%%%
\section{Resolving the plasmon dispersion from the total two-excitation intensity}
\label{PD}
%%%%%%%%%%%%%%%%%%%%%%%%%%%%%%%%%%%%%%%%%%%%%%%%%%%%%%%%%%%%%%%%%%%%%%%%%%%%%%%%%%%%%%%%%%%%%%%%%

In this Section we present one of our main results --- the single-plasmon dispersion as extracted from the total two-excitation intensity by relating various spectral features to important threshold processes. We also establish the framework for interpreting complex spectra in terms of partial contributions.

The distinct features on the total two-excitation intensity curves, shown in Figs. \ref{Fig14}, \ref{Fig15}, \ref{Fig16}, and \ref{Fig17}, provide enough information for extracting the entire single-plasmon dispersion. The position of the low-frequency minimum on two-plasmon curves is given by Eq.~(\ref{OmegaLM}). This expression can be used right away since the dispersion starts at the plasma frequency $\Omega_{pl} = \sqrt{4 \pi n e^2 / m}$. This gives $\omega_{pl}(Q)=\Omega_{lm} - \Omega_{pl}$, where $\Omega_{lm}$ is the minimum position. Note, however, that at $Q > Q_m/2$ the exact position of the low-frequency minimum
on the total curve is no longer determined solely by the $2$-plasmon contribution since it becomes relatively broad and derivatives from the other two processes shift it. Thus, it is best to restrict the plasmon dispersion analysis using this spectral feature to data for $Q \leq Q_m/2$. This is not a problem since the second half of the $[0,Q_m]$ interval can be covered by measuring $2\omega_{pl}(Q/2)$ from the threshold of the two-plasmon spectrum (low-frequency kink shown by the arrow in the lower right panel of Fig.~\ref{Fig15}). This feature is very sharp and provides accurate data all the way to $Q \lesssim 2Q_m$.

One also needs to know the end-point of plasmon dispersion, $\Omega_m=\omega_{pl}(Q_m)$ (from $\Omega_m$ one can easily find $Q_m$ via (\ref{WEH1}) and (\ref{WEH2})). $\Omega_m$ can be found by measuring the intensity around the spectral anomaly at $Q \to 0$, see Figs. \ref{Fig16} and \ref{Fig17}. At $Q=0$ the anomaly is located precisely at $2\Omega_m$. At small but finite $Q$, the anomaly's minimum position is determined by the upper threshold of the hybrid process, $\Omega_m + \Omega_{e-h}(Q_m+Q)$. This then gives access to information about single $e-h$ processes.

An alternative way to measure $Q_m$ and $\Omega_m$ is to look at the $Q \lesssim Q_m$ single-pair process which is not supposed to be masked by the multi-excitation processes. By determining at which momentum $Q$ the single-pair intensity at the maximum starts to decrease, one can locate $Q_m$ and, correspondingly, $\Omega_m$ (see Section \ref{SEP}).

\begin{figure}[tbh]
\centerline{\includegraphics[angle = 0,width=0.8\columnwidth]{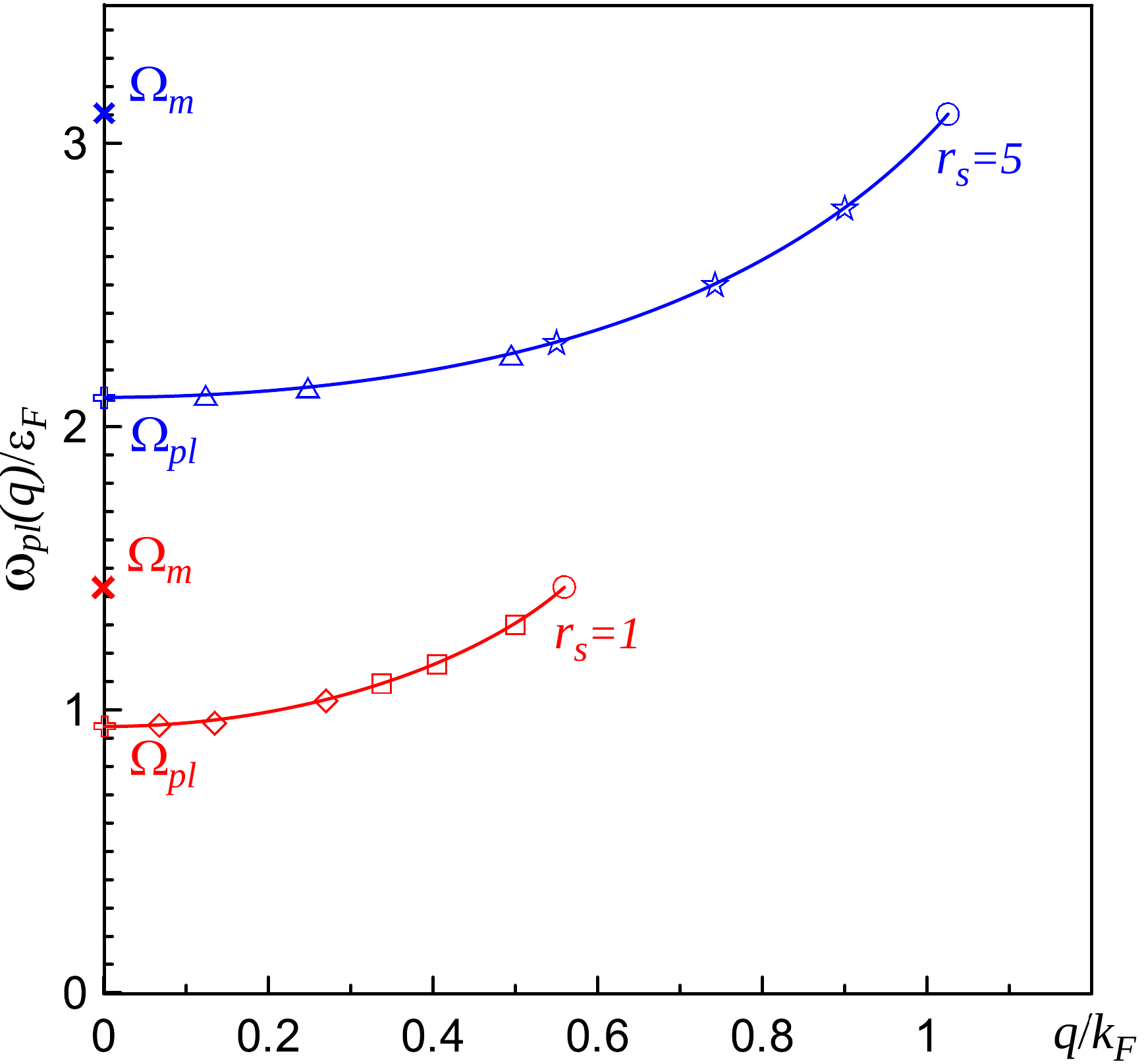}}
\caption{(color online) A comparison of the plasmon dispersions $\omega_{pl}(q)$ for $r_s=1$ (red solid line) and $r_s=5$ (blue solid line) derived from the zeros of the dielectric function with the ones deduced from the two-excitation RIXS spectra (like the ones shown in Figs.~\ref{Fig14}, \ref{Fig15}, \ref{Fig16}). The different symbols correspond to positions of different spectral features. $\Omega_{pl} = \omega_{pl}(0)$ is the plasma frequency and $\Omega_m = \omega_{pl}(Q_m)$ is the maximal plasma frequency. For details see Section \ref{PD}.}
\label{Fig17a}
\end{figure}

With these observation we present the derived plasmon dispersion in Fig.~\ref{Fig17a}. To get the dispersion curves shown in this Figure, we have only used positions of the low-frequency minimum and the low-frequency kink. For $r_s = 1$ we have considered points $Q/k_F = 0.0675$, $0.135$, $0.27$, $0.675$, $0.81$, $1.0$ while for $r_s = 5$ we have used $Q/k_F = 0.124$, $0.248$, $0.495$, $1.10$, $1.485$, $1.80$. The two different sets of symbols on the dispersion curves correspond to applying Eqn. (\ref{OmegaLM}) (the equation governing the low-frequency minimum and used at lower frequencies) and $2 \omega_{pl}(Q/2)$ (the position of the low-frequency kink and used at higher frequencies). Fig.~\ref{Fig17} has been used to establish the value of $\Omega_m$, and thus $Q_m$, see Eq.(\ref{WEH2}). For $r_s=1$ and $5$ we have reproduced $Q_m/k_F = 0.56$ and $Q_m/k_F = 1.027$, respectively ($\Omega_m/\varepsilon_F = 1.433$ and $\Omega_m/\varepsilon_F = 3.109$); in Fig.~\ref{Fig17a} these end points are marked by red and blue circles (red and blue crosses mark the start points, i.e. $\Omega_{pl}(r_s)$).

It is also worth mentioning that, depending on experimental conditions, narrow plasmon peaks may be hard to locate and the entire single-excitation intensity quickly vanishes at small momenta (see, for instance, Fig.~\ref{Fig18} in Section \ref{TOT}).

%%%%%%%%%%%%%%%%%%%%%%%%%%%%%%%%%%%%%%%%%%%%%%%%%%%%%%%%%%%%%%%%%%%%%%%%%%%%%%%%%%%%%%%%%%%%%%%%%%%
\section{First- {\it vs} second-order spectra}
\label{TOT}
%%%%%%%%%%%%%%%%%%%%%%%%%%%%%%%%%%%%%%%%%%%%%%%%%%%%%%%%%%%%%%%%%%%%%%%%%%%%%%%%%%%%%%%%%%%%%%%%%%%

We are now in a position to explore how the first- and second-order processes could appear in experiment for different values of $Q$. For this end we plot the total intensity
\be
I_{12} = D^{(1)} + D^{(2)}/\Gamma^2
\label{I12}
\ee
and consider, for certainty, $(\Omega_{pl}/\Gamma)^2 = 0.1$. [Practically speaking, there is a wide class of actively studied Dirac materials where the chemical potential is close to the Dirac point
and the above condition can be met; in Dirac materials the conduction electrons are typically residing in p-bands.] To present data for the single-plasmon resonance, we smear the $\delta$-functional peak into a Gaussian of half-width $\sigma$ and assume several experimental frequency resolution parameters: $\sigma/\varepsilon_F = 0.01, \; 0.1, \; 0.25,$ and $0.5$. [We do not smear the $e-h$ continuum and spectral densities entering the two-excitation
calculations]. The smallest and the largest values of $\sigma$ are, probably, less realistic,
but need to be considered for completeness of the picture. The most interesting comparison comes from relatively small values of $Q/k_F$ when the two-excitation processes may dominate in the total signal.

\begin{figure}[tbh]
\subfigure{\includegraphics[scale=0.24]{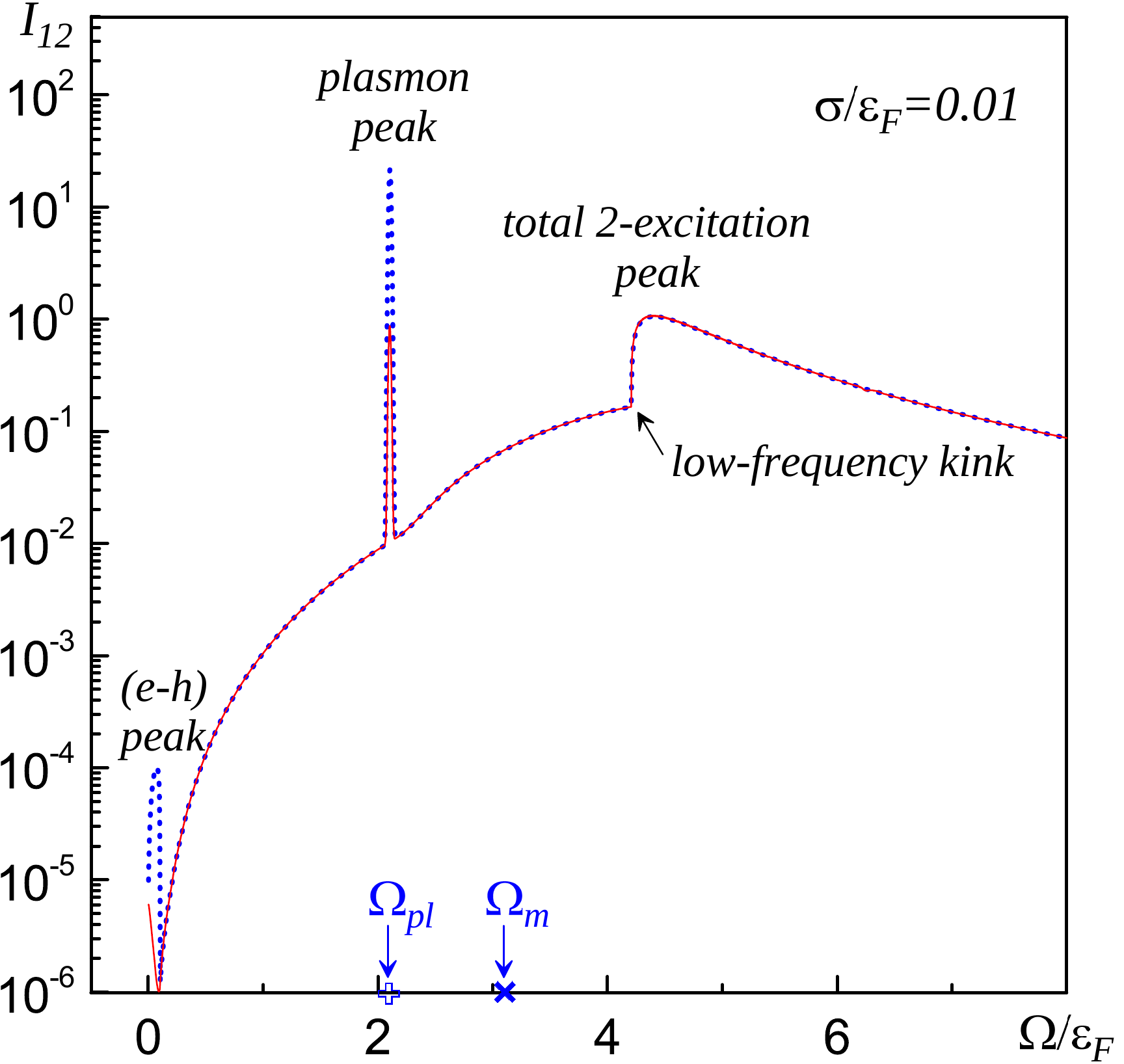}}
\subfigure{\includegraphics[scale=0.24]{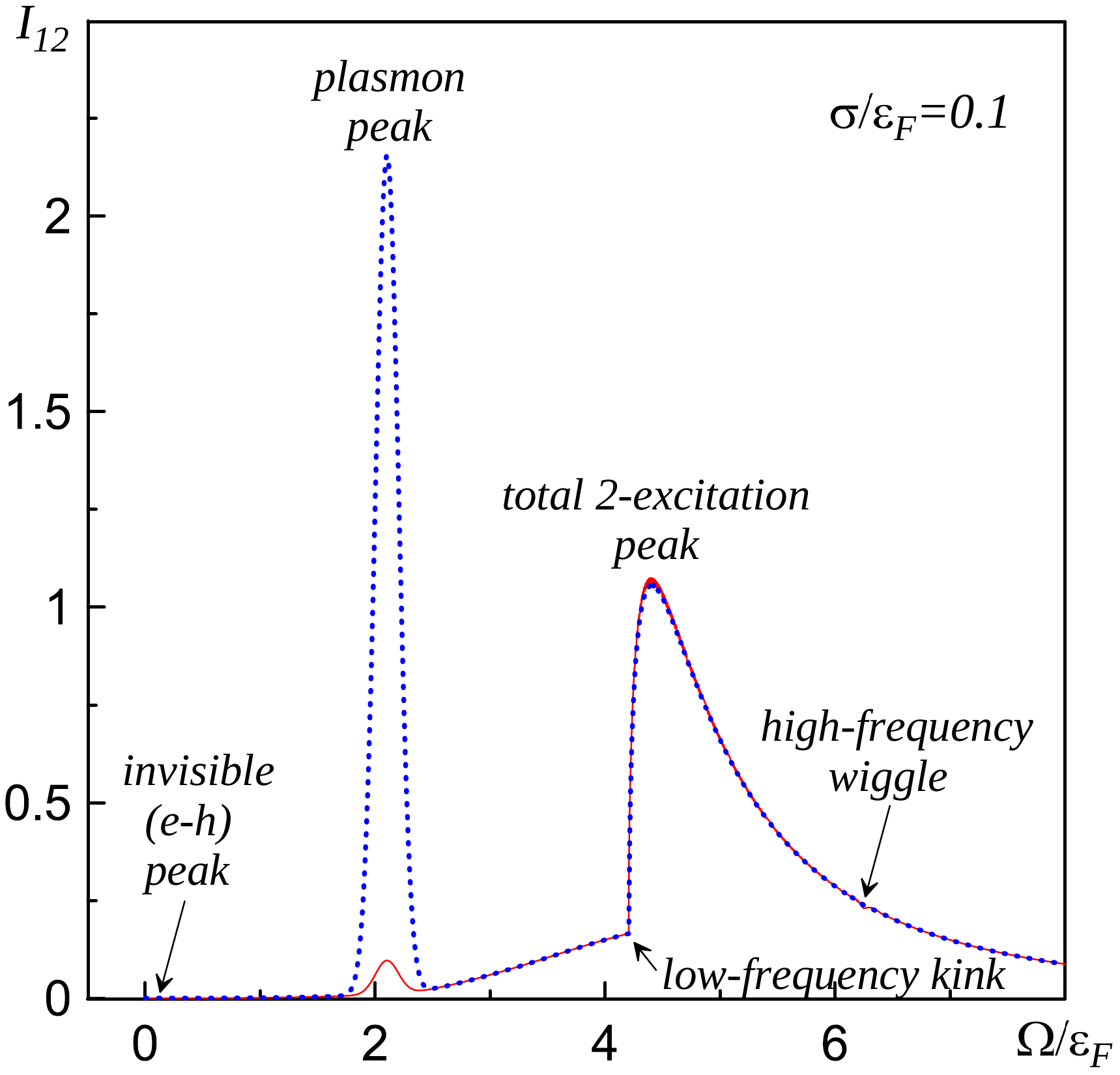}}
\subfigure{\includegraphics[scale=0.24]{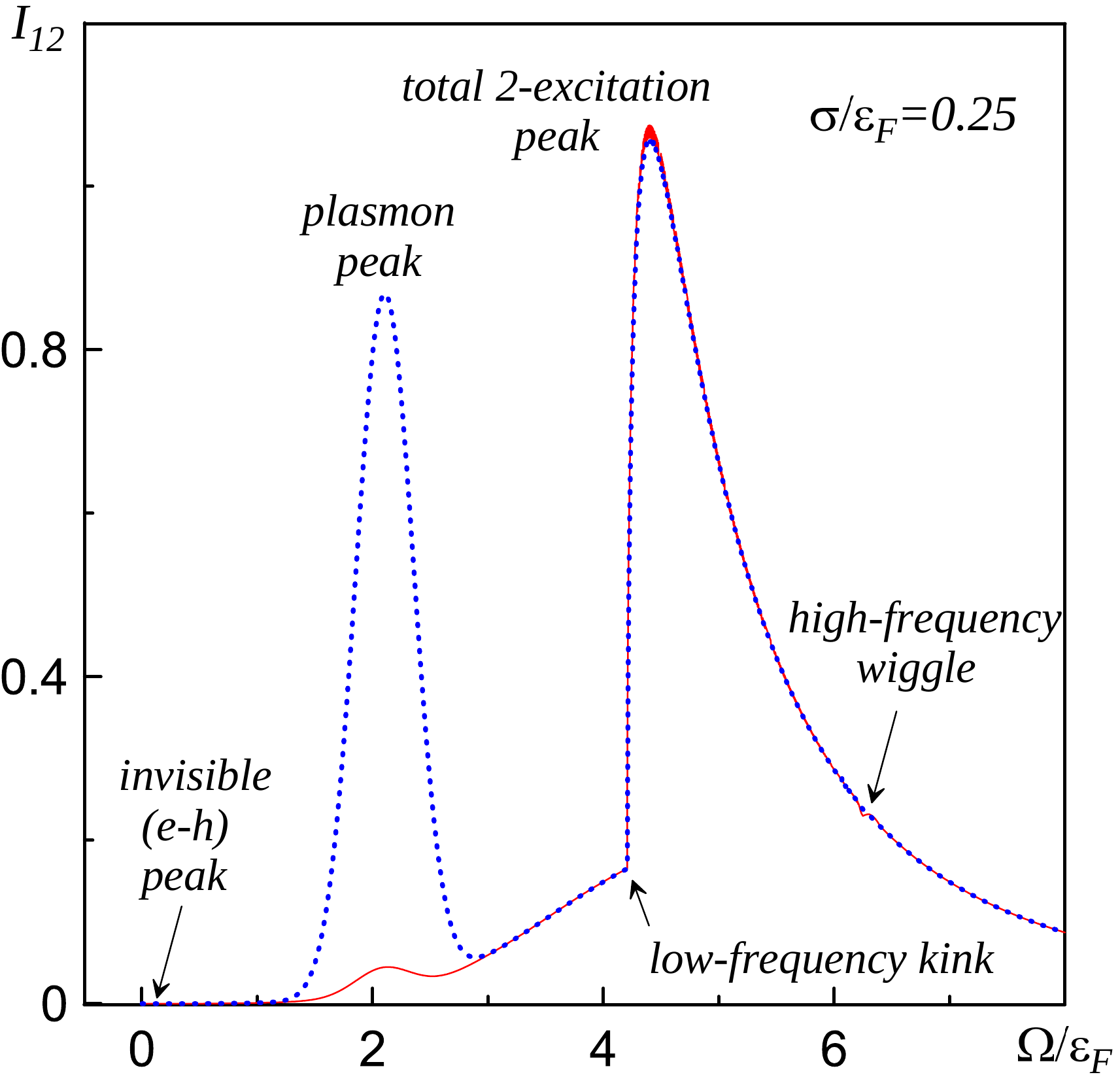}}
\subfigure{\includegraphics[scale=0.24]{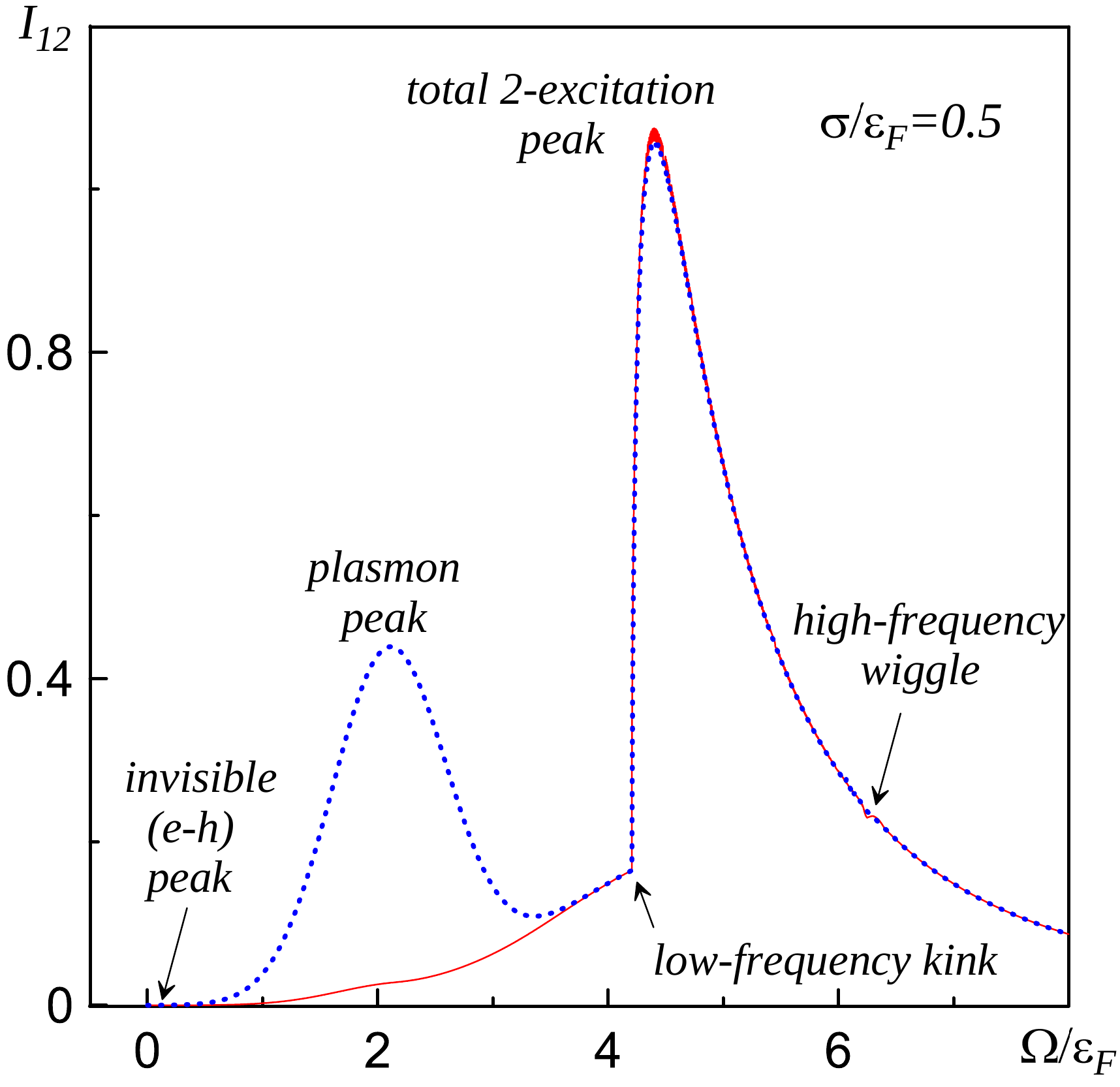}}
\caption{(color online) Total intensities based on Eq.~\ref{I12} as functions of $\Omega$ for small values of momentum transfer $Q/k_F=0.05$ (dotted blue line) and $Q/k_F=0.01$ (red line) at $r_s=5$. The half-width of the smeared single-plasmon peak is increasing from $\sigma/\varepsilon_F$=0.01 in the upper left panel, to $0.1$ in the upper right panel, to 0.25 in the lower left panel, and $0.5$  in the lower right panel. Blue symbols in the upper left panel mark $\Omega_{pl}$ and $\Omega_m$ values for $r_s=5$. }
\label{Fig18}
\end{figure}

In Fig.~\ref{Fig18} we plot the total intensity $I_{12}$ at small momentum transfers
$Q/k_F = 0.01$ and $0.05$ for $r_s=5$. To reveal vastly different intensities associated with
various processes for $\sigma/\varepsilon_F=0.01$ one needs to use the logarithmic scale
in the upper left panel. The upper right and lower left panels allow the reader to
gauge various contributions to intensity from areas under the peaks.
As one can see, at small momentum transfer the single-excitation features, including the plasmon peak, are severely suppressed. The $e-h$ contribution is barely visible in all panels.
However, one can still extract the information about single-particle excitation from the higher-frequency part by measuring various curve anomalies discussed in this work.

\begin{figure}[tbh]
\subfigure{\includegraphics[scale=0.24]{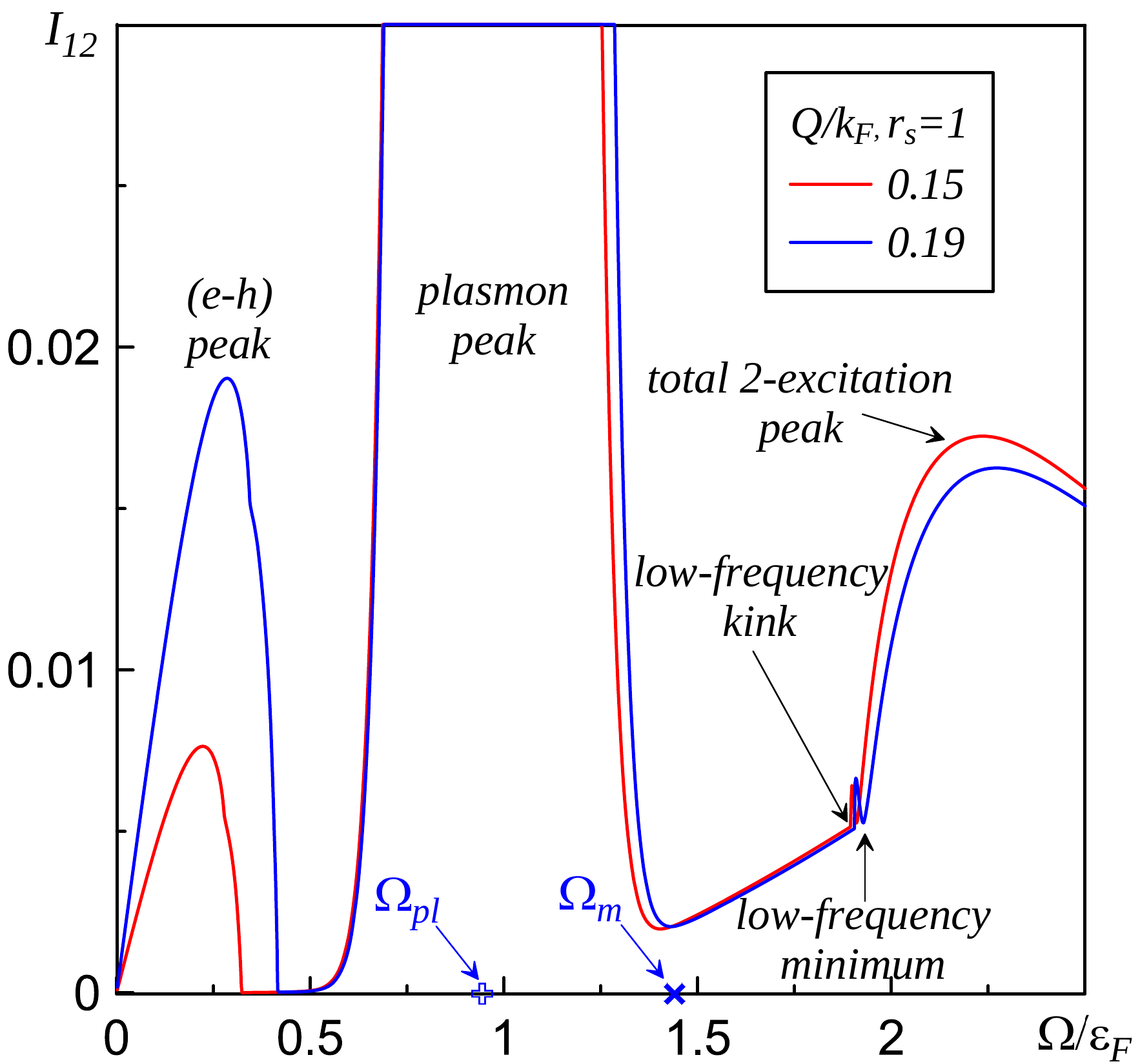}}
\subfigure{\includegraphics[scale=0.24]{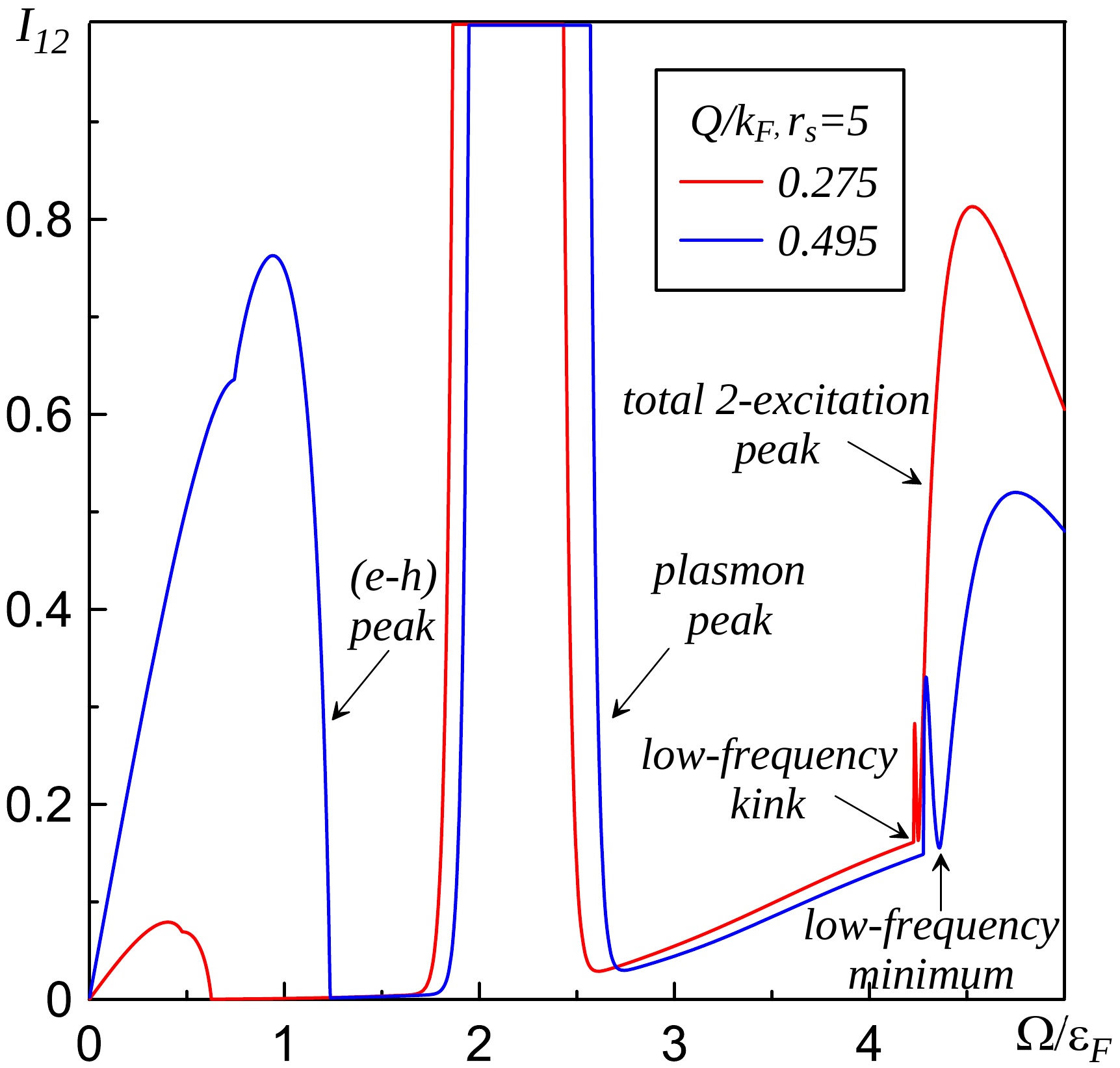}}
\caption{(color online) Total intensities (as in Fig.~\ref{Fig18}) for larger values of
momentum transfer $Q$ at $r_s=1$ (left panel) and $5$ (right panel) for $\sigma/\varepsilon_F=0.1$. Large single-plasmon intensities were cut to fit into the figures. Blue symbols in the left panel mark $\Omega_{pl}$ and $\Omega_m$ for $r_s=1$. }
\label{Fig19}
\end{figure}

In Fig.~\ref{Fig19} we plot the total intensity at $r_s = 1$ and $5$ for larger values of momentum transfer $Q$, when intensities of single-excitation processes become far more prominent while
avoiding substantial overlap with the multi-excitation processes (we take $\sigma/\varepsilon_F=0.1$). Now one can visibly resolve the low-frequency two-plasmon minimum and maximum. Obviously, depending on the system and experimental resolution these spectral features can be also smeared.

%%%%%%%%%%%%%%%%%%%%%%%%%%%%%%%%%%%%%%%%%%%%%%%%%%%%%%%%%%%%%%%%%%%%%%%%%%%%%%%%%%%%%%%%%%%%%%%%%
\section{Conclusions}
\label{CON}
%%%%%%%%%%%%%%%%%%%%%%%%%%%%%%%%%%%%%%%%%%%%%%%%%%%%%%%%%%%%%%%%%%%%%%%%%%%%%%%%%%%%%%%%%%%%%%%%%

We have used the Feynman diagram approach to study the indirect RIXS processes in Coulomb systems in the ultra-short core hole life time approximation often used in the RIXS calculations. We have discussed the single- and two-excitation processes. For the latter we have provided a comprehensive semi-quantitative picture of their contributions. We have demonstrated the need to account for such excitations at small momentum transfer because the single particle contribution here is suppressed by the size of the matrix element. We have further argued that in the limit under consideration, higher-order processes are suppressed and can be neglected.

We have demonstrated that the multi-excitation processes are important from both the fundamental and practical perspectives by showing how to extract the single-plasmon dispersion from the total two-excitation intensity. This can be done by analyzing the universal spectral features of intensity curves revealed in our work. It is worth mentioning that depending on experimental conditions the sharp plasmon resonance at low momenta may be rather difficult to observe because its intensity vanishes in the $Q\to 0$ limit (in contrast to the multi-excitation processes). The intensity vanishing at $Q \to 0$ is also a characteristic feature of the single electron-hole process.

Our analysis is based on the Random Phase Approximation. We believe that this approximation does not qualitatively affect the universal properties of RIXS spectra in metals which originate from thresholds in the particle emission. However, there is no doubt that to obtain quantitatively accurate results for large values of $r_s$, one has to go beyond RPA to account for the renormalization of the Fermi-liquid parameters and vertex corrections. This constitutes the main direction for the future work.

\section{Acknowledgements}

AMT, RMK, and IST thank support from the Office of Basic Energy Sciences, Material Sciences and Engineering Division, U.S. Department of Energy (DOE) under Contract No. DE-SC0012704. NVP thanks support from the Simons Collaboration on the Many Electron Problem.  The authors thank Mark Dean for helpful conversations.

\end{document}